\documentclass[PAPER, atlasdraft=false, cernpreprint=true, texlive=2023, UKenglish,orcidlogo, texmf, orcidlogo]{atlasdoc}
\usepackage{atlaspackage}
\usepackage{atlasbiblatex}
 
\usepackage{atlasphysics}

\graphicspath{{logos/}{figures/}}
 
\usepackage{ANA-STDM-2022-01-PAPER-defs}
 
\usepackage{todonotes}
\DeclareBoolOption[true]{LANGEDIT}
\LEcontact{Chip Brock brockr@msu.edu}

\AtlasTitle{Observation of gauge boson joint-polarisation states in $W^{\pm}Z$ production from $pp$ collisions at $\sqrt{s} = 13$~TeV with the ATLAS detector}

\AtlasAbstract{
Measurements of joint-polarisation states of $W$ and $Z$ gauge bosons in $W^{\pm}Z$ production are presented.
The data set used corresponds to an integrated luminosity of \mbox{$139$~fb$^{-1}$} of proton–proton collisions at a centre-of-mass energy of $13$~TeV recorded by the ATLAS detector at the CERN Large Hadron Collider.
The $W^{\pm}Z$ candidate events are reconstructed using leptonic decay modes of the gauge bosons into electrons and muons.
The simultaneous pair-production of longitudinally polarised vector bosons is measured for the first time with a significance of $7.1$ standard deviations.
The measured joint helicity fractions integrated over the fiducial region are $f_{\mathrm{00}} = 0.067 \pm 0.010$, $f_{\mathrm{0T}} = 0.110 \pm 0.029$, $f_{\mathrm{T0}} = 0.179 \pm 0.023$ and $f_{\mathrm{TT}} = 0.644 \pm 0.032$, in agreement with the next-to-leading-order Standard Model predictions.
Individual helicity fractions of the $W$ and $Z$ bosons are also measured and found to be consistent with joint helicity fractions within
the expected amounts of correlation.
Both the joint and individual helicity fractions are also measured separately in $W^+Z$ and $W^-Z$ events.
Inclusive and differential cross sections for several kinematic observables sensitive to polarisation are presented.

}

\AtlasRefCode{STDM-2022-01}
 
\PreprintIdNumber{CERN-EP-2022-202}

\AtlasJournalRef{\PLB 843 (2023) 137895}
\AtlasDOI{10.1016/j.physletb.2023.137895}

\hypersetup{pdftitle={ATLAS document},pdfauthor={The ATLAS Collaboration}}

\newcommand{\AtlasCoordFootnote}{
ATLAS uses a right-handed coordinate system with its origin at the nominal interaction point (IP)
in the centre of the detector and the \(z\)-axis along the beam pipe.
The \(x\)-axis points from the IP to the centre of the LHC ring,
and the \(y\)-axis points upwards.
Cylindrical coordinates \((r,\phi)\) are used in the transverse plane,
\(\phi\) being the azimuthal angle around the \(z\)-axis.
The pseudorapidity is defined in terms of the polar angle \(\theta\) as \(\eta = -\ln \tan(\theta/2)\).
Angular distance is measured in units of \(\Delta R \equiv \sqrt{(\Delta\eta)^{2} + (\Delta\phi)^{2}}\).}

\DeclareUnicodeCharacter{306}{\v{}}

\usepackage{rotating}
 
\begin{document}
 
\maketitle

\section{Introduction}

 
The study of weak boson polarisation provides a way to directly probe the  Standard Model (SM) gauge and Higgs sectors and to discriminate between SM and beyond the Standard Model (BSM) signals.
The existence of the longitudinally polarised state of weak bosons is a consequence of the non-vanishing mass of the bosons generated by the electroweak (EW) symmetry breaking mechanism.
The measurement of the polarisation in diboson production therefore tests both the SM's gauge symmetry structure, through the existence of the triple gauge coupling, and the particular way this symmetry is spontaneously broken, via the longitudinal helicity state.
Angular observables can be used to look for new interactions that can lead to different polarisation behaviour than predicted by the SM, to which the \wz final state would be particularly sensitive~\cite{Azatov:2017kzw,Panico:2017frx}.

At hadron colliders, the polarisation of the $W$ boson was previously measured in the decay of the top quark
by the CDF and D\O~\cite{Aaltonen:2012rz,Aaltonen:2010ha,Abazov:2010jn} Collaborations and by the ATLAS~\cite{Aad:2012ky} and
CMS~\cite{Khachatryan:2016fky} Collaborations, as well as in single $W$production by ATLAS~\cite{ATLAS:2012au} and CMS~\cite{Chatrchyan:2011ig}.
Polarisation and several other related angular coefficient measurements of singly produced $Z$ bosons were published by the CDF~\cite{Aaltonen:2011nr}, CMS~\cite{Khachatryan:2015paa} and ATLAS~\cite{Aad:2016izn} Collaborations.
The polarisation of $W$ bosons was also measured in $ep$ collisions by the H1 Collaboration~\cite{Aaron:2009wp}.
Finally, for dibosons,  first measurements of the $W$ polarisation were performed by LEP experiments in $W$ pair production in $e^+e^-$ collisions~\cite{Achard:2002bv,Abbiendi:2003wv} and were used to set limits on anomalous triple gauge couplings in Ref.~\cite{Abdallah:2008sf}.
Correlations between polarisation states of $W$ particles were also probed experimentally at LEP~\cite{OPAL:2000wbs,L3:2003tnr,DELPHI:2009wdg}.
Helicity fractions of pair-produced vector bosons were measured independently of each other and for the first time in hadronic collisions at a centre-of-mass energy of $\sqrt{s} = 13$~\TeV\ by the ATLAS Collaboration~\cite{WZPol36fb}, followed by the CMS Collaboration~\cite{CMS:2021icx}, using integrated luminosities of $36~\ifb$ and $137~\ifb$, respectively.
Benefiting from the full ATLAS data set available at $\sqrt{s} = 13$~\TeV, combined with novel analysis techniques,  observables more difficult to extract from data, such as the correlations between helicity states of the vector bosons, can now be accessed.

This Letter reports on an observation and measurement of joint polarisation states of weak bosons in \wz production, extending the pioneering work of Ref.~\cite{WZPol36fb}.
The $W$ and $Z$ bosons are reconstructed using their decay modes into electrons or muons.
The $pp$ collision data were collected with the ATLAS detector from $2015$ to $2018$ at a centre-of-mass energy of $\sqrt{s} = 13$~\TeV\ and correspond to an integrated luminosity of $139~\ifb$.
The individual and joint boson polarisation state fractions are measured consistently in the detector fiducial region and their relationships are studied.
The reported measurements are compared to the recent fixed-order theory calculation of polarised $W$ and $Z$ boson pair production at next-to-leading-order (\NLO) accuracy in quantum chromodynamics (QCD)~\cite{Denner:2020ww,Denner:2020wz}. 
Additionally, the \wz\ production cross section is measured within a fiducial phase space both inclusively and differentially as a function of several angular variables related to the helicity states of the \wz\ system.


\section{Theoretical framework and measurement phase space}

A complete description of the polarisation states of the produced $W$ and $Z$
bosons is given in terms of the two-particle joint spin density matrix
$\rho_{\lambda_W \lambda_W^{'} \lambda_Z \lambda_Z^{'}}$ expressed in the basis
of joint-helicity states $(\lambda_W\lambda_Z)$ where $\lambda_W$ and
$\lambda_Z$ are the helicities of the $W$ and $Z$ bosons, respectively.
In terms of the \wz\ production helicity amplitudes, $F^{(\mu_q \mu_{\bar{q}})}_{\lambda_W \lambda_Z}$, the spin density matrix elements are defined by
\begin{equation}
\rho_{\lambda_W \lambda_W^{'} \lambda_Z \lambda_Z^{'}}  \equiv
\frac{1}{C} \times
\sum_{\mu_q \mu_{\bar{q}}}
F^{(\mu_q \mu_{\bar{q}})}_{\lambda_W \lambda_Z}
F^{(\mu_q \mu_{\bar{q}})*}_{\lambda_W^{'} \lambda_Z^{'}}  \quad \mathrm{, where} \quad
C = \sum_{\mu_q \mu_{\bar{q}} \lambda_W \lambda_Z}
\left| F^{(\mu_q \mu_{\bar{q}})}_{\lambda_W \lambda_Z} \right| ^2 \; . \nonumber
\end{equation}
The initial-state helicity is averaged in the numerator and denominator with
sums running over $\mu_q, \mu_{\bar{q}} = \pm 1/2$.
In the denominator the $W$ and $Z$ bosons helicities are summed over $\lambda_i$, where $\lambda_i = \pm 1,0$, to normalise the trace of the matrix to unity.
The complete $9 \times 9$ joint spin density matrix $\rho_{\lambda_W \lambda_W^{'} \lambda_Z \lambda_Z^{'}}$ cannot yet be measured because of the limited available data sample.
Instead, the following four linear combinations of matrix elements are considered
\begin{equation}
\begin{array}{l l l}
\fzz & = & \rho_{0000} \; , \\
\fTT & = &  \rho_{++--} + \rho_{--++} + \rho_{----} + \rho_{++++} \; , \\
\fzT & = &  \rho_{00--} + \rho_{00++} \; , \\
\fTz & = &  \rho_{--00} + \rho_{++00} \; .\\
\end{array} \nonumber
\end{equation}
The quantities \fzz, \fTT, \fzT and \fTz are composed of diagonal elements of the full matrix and can be interpreted as probabilities of correlated helicity states of the two $W$ and $Z$ bosons.
The distinction between the two transverse helicities of the bosons (right-handed, R, and left-handed, L, also indicated with $+$ and $-$, respectively) is not made.
 
The single-boson spin density matrix $\rho_{\lambda \lambda^{'}}$ is derived from the joint spin density matrix $\rho_{\lambda_W \lambda_W^{'} \lambda_Z \lambda_Z^{'}}$ by summing over indices of one of the bosons.
It provides information about the helicity of one of the bosons regardless of the state of the other.
Diagonal elements of the single-boson spin density matrix are the individual helicity fractions \f0, \fL and \fR for the longitudinal, transverse left-handed and transverse right-handed helicity states, respectively.
The normalisation is defined such that $\f0 + \fL + \fR = 1$.

Helicity fractions are not Lorentz invariant and hence depend on a chosen reference frame.
In diboson production a natural frame choice
for single-boson helicity fraction measurement is to fix the reference $z$-axis
to the direction of the corresponding boson in the diboson centre-of-mass frame~\cite{Denner:2020wz}.
This definition is called the modified helicity coordinate system and is used for this measurement.
The back-to-back kinematics of the bosons in the \wz\ centre-of-mass frame
implies that the polarisation vectors of the two bosons are defined with
respect to the same reference axis, but in opposite directions. 
This choice also maximises the decorrelation between 00 and TT helicity states.

The relative contributions of joint helicity states can be determined from the
absolute value of the cosine of the scattering angle of the $W$ or $Z$  boson
in the $WZ$ centre-of-mass frame, \ctv, calculated with respect to the beam axis~\cite{Baur:1994ia,Franceschini:2017xkh}.
Events with 00 helicity have \ctv values closer to $0$ while events with TT helicity have \ctv values closer to $1$.
 
The polarisation of an individual gauge boson  can be determined from the angular distribution of its decay products.
At the Born level, the expected angular distribution for massless fermions in the rest frame of the parent $W$ boson is given by~\cite{Mirkes:1994eb,Groote:2012xr,Stirling:2012zt,Gounaris:1992kp}
\begin{equation}
\label{eq:WPolarisation}
\frac{1}{\sigma_{W^{\pm}Z}}\frac{\mathrm{d}\sigma_{W^{\pm}Z}}{\mathrm{d}\cos\theta^*_{\ell^W}} = \frac{3}{8}f_{\text{L}}[(1 - q_W \cdot \cos\theta^*_{\ell^W})^2]+\frac{3}{8}f_{\text{R}}[(1 + q_W \cdot \cos\theta^*_{\ell^W})^2]+\frac{3}{4}f_0\sin^2\theta^*_{\ell^W} \; ,
\end{equation}
where $q_W$ is the charge of the $W$ boson and $\theta^*_{\ell^W}$ is defined using the modified helicity frame as the decay angle of the charged lepton in the $W$ rest frame relative to the $W$ direction in the \wz centre-of-mass frame.
All dependencies on the azimuthal angle are integrated over.
The expected angular distribution of the lepton decay products of the $Z$ boson is described by the generalisation of Equation~(\ref{eq:WPolarisation})~\cite{Mirkes:1994eb,Groote:2012xr,Stirling:2012zt}
\begin{equation}
\label{eq:ZPolarisation}
\frac{1}{\sigma_{W^{\pm}Z}}\frac{\mathrm{d}\sigma_{W^{\pm}Z}}{\mathrm{d}\cos\theta^*_{\ell^Z}} = \frac{3}{8}f_{\text{L}}(1+2 \alpha \cos\theta^*_{\ell^Z}+\cos^2\theta^*_{\ell^Z}) + \frac{3}{8}f_{\text{R}}(1 - 2 \alpha  \cos\theta^*_{\ell^Z} + \cos^2\theta^*_{\ell^Z} ) + \frac{3}{4}f_0\sin^2\theta^*_{\ell^Z} \; ,
\end{equation}
where $\theta^*_{\ell^Z}$ is defined using the modified helicity frame as the decay angle of the negatively charged lepton in the $Z$ rest frame relative to the $Z$ direction in the \wz centre-of-mass frame.
The parameter $\alpha =(2c_vc_a)/(c_v^2+c_a^2) = 0.147$ is expressed in terms of the vector $c_v=-\frac{1}{2}+2\sin^2{\theta_{\mathrm{W}}^{\mathrm{eff}}}$ and axial-vector $c_a=-\frac{1}{2}$  couplings of the $Z$ boson to leptons, where the effective value of the Weinberg angle $\sin^2{\theta_{\mathrm{W}}^{\mathrm{eff}}}=0.23152$~\cite{PhysRevD.98.030001} is used.
Equation~(\ref{eq:ZPolarisation}) also holds for the contribution from $\gamma^*$ and its interference with the $Z$ boson, with appropriate $c_v$ and $c_a$ coefficients.

The reported helicity fractions and cross sections are measured in a fiducial phase space chosen to closely follow  the event selection criteria described in Section~\ref{sec:selection}.
It is based on the kinematics of particle-level objects as defined in Ref.~\cite{TruthWS}.
These are final-state prompt\footnote{A prompt lepton is a lepton that is not produced in the decay of a hadron, a $\tau$-lepton, or their descendants.} leptons associated with the $W$ and $Z$ boson decays.
The kinematics of final-state prompt leptons is defined before QED final-state radiation (FSR), i.e.~Born leptons.

The Born charged leptons from $Z$ and $W$ boson decay must have $|\eta| < 2.5$ and transverse momentum \pt\ above $15$~\GeV\ and  $20$~\GeV, respectively;
the invariant mass of the two leptons from the $Z$ boson decay differs by at most $10$~\GeV{}
from the world average value of the $Z$ boson mass $m_Z^{\textrm{PDG}}= 91.1876$~\GeV~\cite{PhysRevD.98.030001}. 
The $W$ transverse mass, defined as $m_{\textrm{T}}^{W} = \sqrt{2 \cdot p_{\mathrm{T}}^\nu \cdot p_{\mathrm{T}}^\ell \cdot[1 -\cos{\Delta \phi(\ell, \nu)}]}$, where $\Delta \phi(\ell, \nu)$
is the angle between the lepton and the neutrino in the transverse plane, and $\pt^\ell$ and $\pt^\nu$ are the transverse momenta of the lepton from $W$ boson decay and of the neutrino, respectively, must be greater than $30$~\GeV.
In addition, it is required that the angular distance $\Delta R$
between the charged leptons from the $W$ and $Z$ decay is larger than $0.3$, and that
$\Delta R$ between the two leptons from the $Z$ decay is larger than $0.2$.
 
The contribution from off-shell effects is minimised by the tight invariant mass window of $\pm 10$~\GeV\ around the nominal $Z$ boson mass used for the definition of the fiducial phase space.
These contributions are estimated from NLO QCD calculations to amount to $2$\% at the integrated level and in most of the differential distributions~\cite{Denner:2020wz}.
However, all the helicity fractions presented here are effective fractions and contain the small non-resonant contributions present in the fiducial phase space, including $\gamma^*$ events.
Interferences among helicity states are also expected to be small, amounting to $0.6$\% according to NLO QCD+EW calculations~\cite{Le:2022lrp}.
Their effect is included in the measured helicity fractions.

A total phase space is also defined by requiring only the invariant mass of the lepton pair associated with the $Z$ boson to be in the range $66 < m_{\ell\ell} < 116$~\GeV.
This serves for the generation of polarised Monte Carlo (MC) event templates described in Section~\ref{sec:Method} for which no restrictions should be applied to the phase space of the decay leptons from $W$ and $Z$ bosons.


\newpage
\section{ATLAS detector}


The ATLAS experiment~\cite{PERF-2007-01} at the LHC is a multipurpose particle detector
with a forward--backward symmetric cylindrical geometry and a near \(4\pi\) coverage in
solid angle.\footnote{\AtlasCoordFootnote}
It consists of an inner tracking detector (ID) surrounded by a thin superconducting solenoid
providing a \SI{2}{\tesla} axial magnetic field, electromagnetic and hadron calorimeters and a muon spectrometer (MS).
The inner tracking detector covers the pseudorapidity range \(|\eta| < 2.5\).
It consists of silicon pixel, silicon microstrip and transition radiation tracking detectors.
Lead/liquid-argon (LAr) sampling calorimeters provide electromagnetic (EM) energy measurements
with high granularity.
A steel/scintillator-tile hadron calorimeter covers the central pseudorapidity range (\(|\eta| < 1.7\)).
The endcap and forward regions are instrumented with LAr calorimeters
for both the EM and hadronic energy measurements up to \(|\eta| = 4.9\).
The muon spectrometer surrounds the calorimeters and is based on
three large superconducting air-core toroidal magnets with eight coils each.
The field integral of the toroids ranges between \num{2.0} and \SI{6.0}{\tesla\metre}
across most of the detector.
The muon spectrometer includes a system of precision tracking chambers and fast detectors for triggering.
A two-level trigger system is used to select events.
The first-level trigger is implemented in hardware and uses a subset of the detector information
to accept events at a rate below \SI{100}{\kHz}.
This is followed by a software-based trigger that
reduces the accepted event rate to \SI{1}{\kHz} on average
depending on the data-taking conditions.
An extensive software suite~\cite{ATL-SOFT-PUB-2021-001} is used in the reconstruction and analysis of real
and simulated data, in detector operations, and in the trigger and data acquisition systems of the experiment.


\section{Signal and background simulation}


MC simulation is used to model signal and most background processes.

An inclusive sample of simulated  \wz\ events is used to correct the signal yield for detector effects and to compare the measurements with the theoretical predictions.
The production of \wz\ pairs and the subsequent leptonic decays of the vector bosons were simulated at NLO in QCD and LO in QED using the
\POWHEG-\textsc{Box}~v2~\cite{powheg1:2004, powheg, Alioli:2010xd, Melia:2011tj} generator,
interfaced to \PYTHIA~8.210~\cite{Sjostrand:2014zea} for simulation of parton showering, hadronisation and the underlying event.
Final-state radiation resulting from QED interactions was simulated using  \PYTHIA~8.210 with the AZNLO~\cite{AZNLO:2014}
set of tuned parameters.
The \CT[10nlo]~\cite{CT10} parton distribution function (PDF) set was used for the hard-scattering process,
while the \textsc{CTEQ6L1}~\cite{Pumplin:2002vw} PDF set was used for the parton shower.
The sample was generated with dynamic renormalisation and factorisation QCD scales,  \muR  and  $\mu_\textrm{f}$, equal to
$m_{WZ}/2$, where $m_{WZ}$ is the invariant mass of the $WZ$ system~\cite{WZPol36fb}.
 
For systematic uncertainties studies, two alternative inclusive \wz signal samples were generated at NLO in QCD.
The first alternative sample was generated using \MGNLO~2.6.5~\cite{Alwall:2014hca}.
Matrix elements containing three leptons, one neutrino and up to two jets in the final state were calculated at NLO QCD and merged with the parton shower from  \PYTHIA~8.210 using the FxFx scheme~\cite{Frederix:2012ps}.  
The \NNPDF[3.0nlo]~\cite{Ball:2014uwa} PDF set was used for the hard-scattering process.
The default dynamic renormalisation and factorisation scale set by \MGNLO was used~\cite{Hirschi:2015iia}.
This MC sample is referred to as \Madgraph.

The second alternative  \wz signal sample was generated using the \Sherpa~2.2.2 generator~\cite{Sherpa:2019gpd}.
Matrix elements contain all diagrams with four electroweak vertices. They were calculated for up to one parton
at NLO and up to three partons at LO using Comix~\cite{Gleisberg:2008fv} and \OPENLOOPS~\cite{Cascioli:2011va},
and merged with the \Sherpa parton shower~\cite{Schumann:2007mg} according to the ME+PS$@$NLO prescription~\cite{Hoeche:2012yf}.
The \NNPDF[3.0nnlo] PDF set was used along with the associated set of tuned parton-shower parameters developed by the \Sherpa authors.

Polarised \wz events were generated at LO in QCD using \MGNLO~2.7.3~\cite{BuarqueFranzosi:2019boy} interfaced to \pythia~8.244 for simulation of parton showering, hadronisation and the underlying event.
The generation of polarised events is not possible at the full \NLO QCD accuracy.
To account for the real part of NLO corrections, events were simulated with either no jets or one jet in the matrix element at LO, and merged with \pythia parton shower using the CKKW-L  scheme~\cite{Lonnblad:2001iq,Lonnblad:2011xx}.
The \NNPDF[3.0nlo] PDF set was used for the hard-scattering process.
Four separate MC samples were generated corresponding to the four joint helicity states  00, 0T, T0 and TT, respectively.
An inclusive MC sample was created by adding the four polarised samples.
These MC sets are referred to as \MG01j.

The background sources in this analysis include processes with two or more electroweak gauge bosons, namely $ZZ$,
$WW$ and $VVV$ ($V=W,Z$); processes with top quarks, such as $t\bar{t}$ and $t\bar{t}V$, single top and \tz; and processes with
gauge bosons associated with jets or photons ($Z+j$ and $Z\gamma$).
Electroweak \wz production, \wzjjew, arising at the order $\alpha_{\mathrm{EW}}^6$ is also considered as a background and not part of the measured signal.
MC simulation is used to estimate the contribution from  background processes with three or more prompt leptons.
Background processes with at least one misidentified lepton are evaluated using data-driven techniques and simulated events are used to assess the systematic uncertainties of these backgrounds.

The \Sherpa~2.2.2 event generator was used to simulate both the $q\bar{q}$ and $gg$-initiated $ZZ$ processes, including $H \rightarrow ZZ$ production, using the \NNPDF[3.0nnlo] PDF set.
It provides a matrix element calculation accurate to NLO in $\alpha_{\mathrm{s}}$ for 0- and 1-jet final states, and LO accuracy for 2- and 3-jet final states.
Triboson events were simulated by the \Sherpa~2.2.2 event generator at NLO accuracy with 0 additional partons and at LO accuracy with one and two additional partons and using the \NNPDF[3.0nnlo]  PDF set.
 
Both the $t\bar{t}Z$ and $t\bar{t}W$ processes were generated at NLO with the \MGNLO MC generator, using the \NNPDF[3.0nlo] PDF set, and interfaced to the \PYTHIA~8.186~\cite{Sjostrand:2007gs} parton shower model.
 
Finally, the \tzj and \wzjjew processes were generated at LO by the  \MGNLO MC generator interfaced with \pythia~8.244 for
the modelling of the parton shower, hadronisation and underlying event.
The parton distribution function set was \NNPDF[3.0nlo].
In the simulation of the parton shower the  \NNPDF[2.3lo]~\cite{Ball:2012cx}  PDF set was used.

All generated MC events were passed through the ATLAS detector simulation~\cite{Aad:2010ah}, based on
\GEANT~\cite{Agostinelli:2002hh}, and processed using the same reconstruction software as used for the data.
The event samples include the simulation of additional $pp$ interactions (pile-up)
generated with \PYTHIA~8.186 using the \NNPDF[2.3lo] PDF set
and the A3~\cite{ATL-PHYS-PUB-2016-017} set of tuned parameters for the underlying event and parton fragmentation.
Simulated events were reweighted to match the pile-up conditions observed in the data.
Scale factors were applied to simulated events to correct for small differences in the efficiencies observed in data and predicted by MC simulation for the trigger, reconstruction, identification, isolation and impact parameter requirements of electrons and muons~\cite{ATLAS:2019qmc,ATLAS:2020auj}.
Furthermore, the electron energy and muon momentum in simulated events were smeared to account for small
differences in resolution between data and MC events~\cite{ATLAS:2019qmc,ATLAS:2016lqx}.


\section{Event reconstruction and selection}
\label{sec:selection}


Only data recorded with stable beam conditions and with all relevant detector subsystems operational are considered.
Candidate events are selected using triggers~\cite{Aaboud:2016leb,ATLAS:2019dpa,ATLAS:2020gty} that require at least one electron or muon.
These triggers require leptons to pass different transverse momentum threshold and isolation criteria, which depend on the data-taking run period and the instantaneous luminosity.
The combined efficiency of these triggers is $99\%$ for \wz events satisfying the offline selection criteria.
Events are required to have a primary vertex compatible with the LHC luminous region inside the ATLAS detector.
The primary vertex is defined as the reconstructed vertex with at least two charged-particle tracks that has the largest sum of the $\pT^{2}$ for the associated tracks.

Muon candidates are identified by tracks reconstructed in the MS and matched to tracks
reconstructed in the ID. Muons are required to pass a `medium' identification selection,
which is based on requirements on the number of hits in the ID and the MS~\cite{ATLAS:2016lqx,ATLAS:2020auj}.
The efficiency of this selection averaged over \pt and $\eta$ is larger than $98\%$.
The muon momentum is measured by combining the MS measurement, corrected for the energy deposited in
the calorimeters, and the ID measurement.
The $\pt$ of the muon must be greater than $15$~\GeV{} and its pseudorapidity must satisfy $|\eta|<2.5$.
 
Electron candidates are reconstructed from energy clusters in the electromagnetic calorimeter matched
to ID tracks. Electrons are identified using a discriminant that is the value of a likelihood
function constructed with information about the shape of the electromagnetic showers in the calorimeter,
the track properties and the quality of the  track-to-cluster matching for the candidate~\cite{ATLAS:2019qmc}.
Electrons must satisfy a `medium' likelihood requirement, which provides an overall identification efficiency of $90\%$.
The electron momentum is computed from the cluster energy and the direction of the track. The $\pt$ of
the electron must be greater than $15$~\GeV{} and the pseudorapidity of the cluster must satisfy
$\abseta< 1.37$ or $1.52 < \abseta < 2.47$.
 
Electron and muon candidates are required to originate from the primary vertex and to be isolated from other particles using both calorimeter-cluster and ID-track information.
The isolation efficiency is at least $90\%$ for electrons with $\pT>25$~\GeV\ and muons with $\pT>15$~\GeV\, and at least $99\%$ for electrons and muons with $\pT>60$~\GeV~\cite{ATLAS:2019qmc,ATLAS:2020auj}.

Jets of hadrons are reconstructed using a particle-flow algorithm based on noise-suppressed positive-energy topological clusters in the calorimeter~\cite{ATLAS:2017ghe}.
Energy deposited in the calorimeter by charged particles is subtracted and replaced by the momenta of tracks which are matched to those topological clusters.
The jets are clustered using the anti-$k_{t}$ algorithm~\cite{antikt,Fastjet} with a
radius parameter $R = 0.4$.
They are  calibrated according to in situ measurements of the jet energy scale~\cite{ATLAS:2020cli}.
All jets must have $\pt>25$~\GeV{} and be reconstructed in the pseudorapidity range $|\eta|<4.5$.
Jets with $\pt$ below $60$~\GeV\ and with $|\eta|<2.4$ have to pass a requirement on the
{\textit{jet vertex tagger}}~\cite{Aad:2015ina}, a likelihood discriminant that uses a combination of
track and vertex information to suppress jets originating from pile-up activity.
Jets with $|\eta|<2.5$ containing a $b$-hadron are identified with a deep-learning neural network (NN)~\cite{ATLAS:2019bwq,ATL-PHYS-PUB-2017-003,ATL-PHYS-PUB-2017-013} which uses distinctive features of $b$-hadron decays in terms of the impact parameters of the tracks and the displaced vertices reconstructed in the inner detector.
Jets initiated by $b$-quarks are selected by setting the algorithm's output threshold such that an $85\%$ $b$-jet selection efficiency is achieved in simulated $t\bar{t}$ events, with a rejection factor of 40 against light-flavour
jets.

The transverse momentum of the neutrino is estimated from the missing transverse momentum in the event, \met,
calculated as the negative vector sum of the transverse momentum of all identified hard physics objects
(electrons, muons, jets), with a contribution from an additional soft term.
This soft term is calculated from ID tracks matched to the primary vertex and not assigned to any of the hard objects~\cite{Aaboud:2018tkc}.

Events are required to contain exactly three lepton candidates satisfying the selection criteria described above.
To ensure that the trigger efficiency is well determined, at least one of the candidate leptons is required to have
$\pt > 25$~\GeV\ for $2015$ and $\pt > 27$~\GeV\ for $2016$--$2018$ data, as well as being  geometrically matched to a lepton that was selected by the trigger.
No requirement on the number of jets is applied.
 
In order to suppress background processes with at least four prompt leptons, events with a fourth lepton candidate
satisfying looser selection criteria are rejected. For this looser selection, the lepton $\pt$ requirement is
lowered to $\pt>5$~\GeV{}, electrons are allowed to be reconstructed in the region $1.37< \abseta< 1.52$ and `loose' identification requirements~\cite{ATLAS:2019qmc,ATLAS:2020auj} are used for both the electrons and muons.
A less stringent requirement is applied for electron isolation and is based only on ID track information.
No dedicated identification algorithm is used to suppress events with electrons and muons originating from the decay of $\tau$-leptons.
 
Candidate events are required to have at least one pair of leptons with the same flavour and opposite charge,
with an invariant mass that is consistent with $m_Z^{\textrm{PDG}}$ to within $10$~\gev.
This pair is considered to be the $Z$ boson candidate. If more than one pair can be formed, the pair whose invariant mass
is closest to the nominal $Z$ boson mass is taken as the \Zboson\ boson candidate.
The remaining third lepton is assigned to the $W$ boson decay.
The transverse mass of the $W$ candidate, computed using
\met and the \pt of the associated lepton, is required to be greater than $30$~\GeV.
 
Backgrounds originating from misidentified leptons are suppressed by requiring the lepton associated with the $W$ boson
to satisfy more stringent selection criteria. The transverse momentum of this lepton is therefore required to be greater than $20$~\GeV. This lepton is also required to pass the `tight'
identification requirements, which results in an efficiency between $90\%$ and $98\%$ for muons and
an overall efficiency of $85\%$ for electrons.

The main challenge in the reconstruction of the full kinematics of the selected events arises for the $W$ boson from incomplete knowledge of the neutrino momentum.
In the previous ATLAS publication~\cite{WZPol36fb} the  neutrino longitudinal momentum $p_z^\nu$ was reconstructed analytically using the $W$ mass constraint.
In this analysis a neural network regression is developed to reconstruct $p_z^\nu$.
Input variables to the NN are  \pt and \pz of the charged lepton from the $W$ boson decay, the \etmiss components perpendicular and parallel to the direction of this charged lepton and the \pznu reconstructed analytically. 
The NN is trained using \wz \powhegpythia MC events at detector level to reconstruct event-by-event the true \pznu value in the MC simulation.
The compatibility of the reconstructed $W$ boson mass \mw with its generated value is ensured by using as the loss function the sum of the root-mean-square (RMS) of the deviations of  \mw and $p_z^\nu$ reconstructed values from their generated values. 
The regression method provides a reasonable estimate of $p_z^\nu$ for events for which the analytical method fails~\footnote{The analytical method relies on a second-order equation which has complex solutions for roughly $30\%$ of the events. In those cases, its imaginary part is set to zero.} and reduces the $p_z^\nu$ error RMS by 10\%
thanks to the improved resolution at high $p_z^\nu$.
This improvement comes from fact that the NN is trained to account for the full width of the generated $W$ boson, whereas the analytical method imposes a single value of the $W$ boson mass,
fixed to its pole mass, for all events.
Additionally, the regression method improves the \mw resolution from $41\%$ obtained with the analytical method to $31\%$.
This in turn improves the reconstruction of angular observables in the modified helicity frame.


\section{Background estimation}

The background sources are classified into two groups: events where at least one of the candidate leptons is not a prompt lepton (reducible background)
and events where all candidates are prompt leptons or are produced in the decay of a $\tau$-lepton (irreducible background). Candidates that are not prompt leptons are also called  `misidentified' leptons.
 
Events in the first group mostly originate from  $Z+j$, $Z\gamma$, $t\bar{t}$, and $WW$ production processes and constitute about $5\%$ of all selected events.
This reducible background is  estimated with a data-driven method based on the inversion of a matrix containing the efficiencies and  misidentification probabilities for prompt and misidentified leptons~\cite{ATLASWZ_8TeV,Aad:2014pda,WZPol36fb,ATLAS:2022swp}. 
The method exploits  the classification of the leptons as  loose  or tight  candidates and  the probability that a non-prompt lepton is misidentified as a loose or tight lepton.
Tight leptons are leptons passing the selection criteria described in Section~\ref{sec:selection}.
Loose leptons are leptons that do not meet the isolation and identification criteria for signal leptons but satisfy only looser criteria.
The misidentification probabilities for misidentified leptons are determined from data using dedicated control samples each enriched  in misidentified leptons from light- or heavy-flavour jets and from photon conversions, respectively.
The lepton efficiencies for prompt leptons are determined as detailed in Refs.~\cite{ATLAS:2019qmc,ATLAS:2016lqx,ATLAS:2020auj}.
The lepton efficiencies and misidentification probabilities are combined with event rates in data samples of \wz candidate events where at least one and up to three of the leptons are loose.
Then, solving a system of linear equations, the number of events with at least one misidentified lepton  is obtained, which represents the amount of reducible background in the \wz sample.
About $2\%$ of this background contribution arises from events with two misidentified leptons.
The background from events with three misidentified leptons, e.g. from multijet processes, is negligible.
The method allows the shape of any kinematic distribution of reducible background events to be estimated.
Another independent method of assessing the reducible background was also considered.
This method estimates the amount of reducible background using MC simulations scaled to data by process-dependent factors determined from the data-to-MC comparison in dedicated control regions.
Agreement within $20\%$ with the matrix method estimate is obtained in both yield and shape of the distributions of irreducible background events.

The events contributing to the second group of background processes represent $\sim18$\% of all selected events.
They originate from $ZZ$, $t\bar{t} +V$, $VVV$ (where $V$ = $Z$ or $W$) and  \tz events.
Events from \wz production with at least one $\tau$-lepton decay or from electroweak \wzjjew production are also considered as backgrounds and not part of the measured signal.
The amount of irreducible  background is estimated using MC simulations. 
The dominant contribution in this second group is from $ZZ$ production, where one of the leptons from the $ZZ$ decay falls outside the detector acceptance. It represents about $7.5\%$ of all selected events.
The MC-based estimates of the $ZZ$ and $t\bar{t} +V$ backgrounds are validated by comparing the MC expectations with the event yield and several kinematic distributions of data samples enriched in $ZZ$ and $t\bar{t} +V $ events, respectively.
The $ZZ$ control sample is selected from events with a $Z$ candidate which meets all of the analysis selection criteria and which is accompanied by two additional leptons, satisfying the lepton criteria described in Section~\ref{sec:selection}.
The $ZZ$ MC expectation is rescaled to match the data by combining  the \wz\ signal region and a single bin containing all of the events from the $ZZ$ control region in the fit detailed in Section~\ref{sec:Method} and used to extract the helicity fractions.
A scaling factor of $1.13$ is obtained from the fit for the $ZZ$ MC contribution, in agreement with the $ZZ$ cross-section measurements performed at $\sqrt{s} = 13$~\TeV~\cite{Aaboud:2017rwm}.
The $t\bar{t} +V$ control sample is selected by requiring selected \wz\ events to have at least two reconstructed $b$-jets.
The observed data event yield in this validation region is matching the $t\bar{t} +V$ MC prediction rescaled by a factor of $1.3$.
This scaling factor relative to predictions is in line with the $t\bar{t} + Z$ cross-section measurements performed at $\sqrt{s} = 13$~\TeV~\cite{ATLAS:2021fzm}. 
The contribution of $t\bar{t} +V$ events amounts to $\sim 4\%$ of all selected events.
Systematic uncertainties affecting the rescaling of $ZZ$ and $t\bar{t} +V $ MC predictions are discussed in Section~\ref{sec:systematics}.
Finally, some $W^\pm \gamma^*$ events produced outside of the total phase space can be selected at detector level.
The contribution of such events is estimated from the \wz \powhegpythia MC simulation to amount to $0.5$\% and is subtracted from data.


\section{Measurement methodology}
\label{sec:Method}

 
\subsection{Joint boson polarisation}

In the Born-level fiducial phase space, the \ctv, \qctw and \ctz angular observables can be used together to separate the joint helicity states from an inclusive MC sample and determine their respective generated fractions.
However, when using the same angular observables at detector level, the separation of the different joint helicity states is not sufficient for the measurement of their respective fractions with an expected precision of 5~$\sigma$.
Therefore, a multivariate discriminant exploiting more kinematic observables  is used to better separate the four joint helicity states.
A deep neural network (DNN) classifier is implemented to exploit kinematic differences between polarisation states in  different observables.
Keras v2.2.4-tf~\cite{chollet2015keras} with the Tensorflow v1.3~\cite{tensorflow2015-whitepaper} backend are used.
The DNN is trained and optimised using simulated \MG01j polarised MC samples.
A total of eight variables are used as inputs to the DNN: the transverse momenta of the three leptons and of the neutrino, $p_\mathrm{T}^{\ell^W}$, $p_\mathrm{T}^{\ell^Z_1}$, $p_\mathrm{T}^{\ell^Z_2}$ and $E_\mathrm{T}^\mathrm{miss}$, angular variables as the absolute difference between the rapidities of the $Z$ boson and the lepton from the decay of the $W$ boson, $|y_Z - y_{\ell^W}|$, the azimuthal angle difference between the two leptons of each $W$ and $Z$-boson decay, $\Delta \phi(\ell^{W},\nu)$, $\Delta \phi(\ell_1^{Z},\ell_2^{Z})$, respectively, and the transverse momentum of the $WZ$ system $p_\mathrm{T}^{WZ}$.
The DNN classifier is trained such that the components of the vector output, one for each of the four joint-polarisation states, are combined in a single output node in its last layer, used in the analysis for the separation.
The optimisation of the combination is part of the DNN training procedure and a sigmoid activation function is applied to the combination output.
The output is therefore a continuous score, labelled `\PolDNNscore,'  ranging from 0 to 1 that takes typical values peaking around 0, 0.5, and 1 for TT, mixed 0T and T0, and 00 helicity states, respectively.
The DNN classifier achieves integrals of the receiver operating characteristic (ROC) curves between 0.8 and 0.85 for separating the 00 state from each of the three others joint polarisation states.
The sensitivity of the analysis in disentangling  the mixed helicity states 0T and T0 is achieved by categorising events depending on the \ctw and \ctz observables.
Four categories are defined as $(\absctw < 0.5,\absctz < 0.5)$, $(\absctw > 0.5,\absctz < 0.5)$, $(\absctw < 0.5,\absctz > 0.5)$ and $(\absctw > 0.5,\absctz > 0.5)$.
A one-dimensional representation of the \PolDNNscore variable in the four $(\absctw,\absctz)$ categories, shown in Figure~\ref{fig:2VResults:postFitDist} and referred to as the \PolCatDNNscore, is built.

To measure joint helicity fractions a binned maximum-likelihood fit~\cite{Cranmer:1456844} to data of the four helicity states templates is performed.
The binned likelihood function used in the fit consists of a product of Poisson probability terms over bins of the \PolCatDNNscore distribution in the \wz\ signal region and a single-bin integrating over the $ZZ$ control region.
Each source of systematic uncertainty is implemented in the likelihood function as a nuisance parameter with a Gaussian constraint, except for the NLO QCD modelling systematic uncertainty described in section~\ref{sec:systematics}.
Independent parameters of the fit are the particle level helicity fractions \fzz, \fTz, \fTT and the integrated fiducial cross section.
The fourth joint helicity fraction \fzT is obtained using the constraint $\fzz + \fzT + \fTz + \fTT = 1$.
Differences between detector-level and particle-level fractions due to detector efficiencies and QED final-state radiation effects are accounted for in the fit using independent correction factors applied to each helicity fraction.
The factors are obtained from simulation and range from 0.95 to 1.01 depending on the helicity state.

As observed in theory calculations~\cite{Denner:2020wz}, a strong relationship exists between \NLO QCD corrections and polarisation effects.
Therefore, templates for helicity states generated at LO are insufficient.
The \MG01j MC simulation corrects for the real part of \NLO QCD effects but misses virtual corrections that are also important for polarisation measurements.
In order to verify that the shapes of the polarised templates are valid, a closure test is performed.
Templates are used to fit pseudo-data generated using inclusive MC simulations at \NLO QCD accuracy, such as with \powhegpythia or \Madgraph.
The fit is performed on detector-level distributions.
Polarisation fractions extracted from the fit are compared with the generated polarisation fractions of the \NLO MC samples. 
Differences from $10$\% to $50$\%  depending on the polarisation state are observed between extracted and built-in fractions when using polarised templates from the \MG01j generation.
This demonstrates the need for using polarised templates at \NLO QCD accuracy for polarisation measurements.

Templates better approaching the \NLO QCD accuracy are built using a DNN-based event-by-event reweighting procedure~\cite{Andreassen:2019nnm}.
Four DNNs are trained and each of them is specialised to reweight at particle-level the inclusive \MG01j events to one of the four joint polarised states.
Input variables of the DNNs are those of the polarisation DNN  classifier augmented by the invariant mass of the $WZ$ system \mwz, and the angular variables \ctw, \ctz and \ctv.
The four DNNs are then applied to reweight the inclusive \powhegpythia MC events, creating four polarised MC templates with NLO-like accuracy in QCD.
This method provides the best fit closure, with no bias on the extracted polarisation fractions visible within the statistical precision of the closure test.

A second method is used to create \NLO QCD accurate polarised templates, based on the available fixed-order parton-level theory predictions~\cite{Denner:2020wz}.
Predicted distributions, including the output score of the DNN classifier, are calculated  in the parton-level fiducial phase space~\cite{Denner:2020wz}. 
Corrections for parton-shower (PS) effects on the shape of the templates are incorporated by binned factors.
The correction factors are calculated per helicity state by comparing \MG01j predictions at particle and parton level.
The templates obtained at particle level are transformed to detector level using a folding procedure to account for detector effects.
This method provides a good fit closure, reducing the bias up to $\sim 10$\%.
This template creation method is used to estimate systematic uncertainties in the template shapes.

Interference among polarisation states manifests itself as a shape difference between the sum of polarised templates and the inclusive distribution.
This difference is expected to be at most $4$\% for the \PolDNNscore observable~\cite{Denner:2020wz}.
The fit procedure adjusts fitted polarisation fractions such that the sum of polarised templates matches the inclusive distribution of data, with increased fit uncertainties.
The measured polarisation fractions therefore include interference effects.

\subsection{Individual boson polarisation}

For individual $W$ and $Z$ boson polarisation states, the helicity parameters \f0 and \fLR are measured using a fit of templates for the three \f0, \fL and \fR helicity states to the \qctw and \ctz  distributions.
The equation $\f0 + \fR + \fL = 1$ is used to constrain the independent parameters of the fit to determine \f0, \fLR and the integrated fiducial cross section.
Templates with \NLO QCD accuracy are also required  for the \qctw and \ctz distributions.
Neither such templates at \NLO QCD accuracy nor templates for left- and right-handed helicity states of the  vector bosons can be produced by the \MGNLO~2.7.3 MC event generator.
The templates for the \qctw and \ctz distributions for each of the three individual helicity states of the $W$ and $Z$ bosons are therefore created from the inclusive \powhegpythia MC sample using a reweighting technique~\cite{WZPol36fb}.
For each of the gauge bosons the predicted helicity fractions of \powhegpythia MC events are determined using the moment method~\cite{Mirkes:1994eb, Aad:2016izn} in the total phase space.
Following the method detailed in  Ref.~\cite{WZPol36fb}, the predicted fractions are used to reweight \powhegpythia MC events to represent each of the longitudinal, left- and right-handed states of the $W$ or $Z$ bosons.
This creates polarised MC templates at detector level.

\subsection{Differential cross section}
 
Independently of polarisation fractions, the inclusive \wz production cross section is measured differentially  within the fiducial phase space defined at particle level and as a function of the variables the most sensitive to polarisation of the vector bosons.
Similarly to measurements in Ref.~\cite{WZPol36fb}, the differential detector-level distributions of background-subtracted data are corrected for detector resolution, efficiencies and QED final-state radiation effects  using simulated signal events and an iterative Bayesian unfolding method~\cite{DAgostini:1994zf}, as implemented in the RooUnfold toolkit~\cite{RooUnfold}.
The number of iterations ranges from two to four, depending on the resolution in the unfolded variable and chosen in order to minimise the sum of the expected statistical and unfolding uncertainties in the unfolded distribution.
The width of the bins in each distribution is chosen according to the experimental resolution and to the statistical significance of the expected number of events in each bin.
The fraction of signal MC events generated in a bin which are reconstructed in the same bin is always greater than 35\%.
The detector-level distributions are unfolded with a response matrix computed using the \powhegpythia MC signal sample that includes all four measured decay channels ($eee$, $ee \mu$, $\mu\mu e$ and $\mu\mu\mu$) and is divided by four, such that cross sections refer to final states where the $W$ and $Z$ bosons decay in a single leptonic channel with muons or electrons.


\section{Systematic uncertainties}
\label{sec:systematics}


Systematic uncertainties  affecting the shape and normalisation of the \PolCatDNNscore distribution for the background contributions, as well as the acceptance of the helicity components and the shape of their templates, are considered.

Uncertainties due to theoretical modelling in the helicity templates for higher order QCD corrections are evaluated by varying in the \powhegpythia MC simulation the renormalisation and factorisation scales independently by factors of two and one-half, removing combinations where the variations differ by a factor of four.
The uncertainties due to the PDF and the $\alphas$ value used in the PDF determination are evaluated using the \texttt{PDF4LHC} prescription~\cite{Butterworth:2015oua}.

For joint polarisation measurements an uncertainty due to the modelling of NLO QCD is estimated by comparing predictions of the \PolCatDNNscore distributions for each of the joint-helicity states obtained from the parton-level \NLO calculation with the DNN reweighting of the \powhegpythia MC simulation.
The difference between the predicted shapes from the two methods is considered as an uncertainty.
The associated nuisance parameter in the fit is parametrised by a flat constraint in the range between the two template shapes from each of the method, with steeply vanishing constraints outside of this range, defined using a double Fermi distribution.
A specific uncertainty is attributed to the DNN reweighting method.
It is estimated by applying the reweighting method to inclusive \MG01j MC events and comparing the obtained distributions with those of polarised \MG01j MC events.
The difference between the shapes is treated as an uncertainty.

For individual boson polarisation measurements a modelling uncertainty in the helicity templates 
is estimated by comparing predictions of the \qctw and \ctz distributions from the \powhegpythia event generator with those of \sherpa.
The difference between the predicted shapes from the two generators is considered as an uncertainty.
An uncertainty is attributed to the reweighting method used to create templates with individually polarised bosons from inclusive \powhegpythia MC events.
It is estimated in the same way as for the joint polarisation measurement.

Uncertainties in differential cross-section measurements arising from imperfect description of the data by the \powhegpythia MC simulation are evaluated using a data-driven method~\cite{Malaescu:2011yg}.
Each MC differential distribution is corrected to match the data distribution and the resulting weighted MC distribution at detector level is unfolded
with the response matrix used in the actual data unfolding.
The new unfolded distribution is compared with the weighted MC distribution at particle level and the difference is taken as the systematic uncertainty.
An additional uncertainty is derived to account for more subtle differences between the \powhegpythia and \Madgraph generators (e.g.\ hadronisation models, additional soft objects, mis-modelling  in other kinematic variables). 
The \Madgraph generator is used to unfold the data and deviation from the nominal result is taken as the uncertainty.
In order to remove effects already accounted for in the data-driven method, the \Madgraph distributions were first reweighed to match \powhegpythia distributions at particle level.

Systematic uncertainties affecting the reconstruction and energy calibration of electrons, muons and jets are propagated through the analysis.
The uncertainties due to lepton reconstruction, identification, isolation requirements and trigger efficiencies as well as in the lepton momentum scale and resolution are assessed using tag-and-probe methods in $Z \to \ell \ell$ events~\cite{ATLAS:2019qmc,ATLAS:2016lqx,ATLAS:2020auj}.
The uncertainties in the jet energy scale and resolution are based on their respective measurements in data~\cite{Aaboud:2017jcu}.
The uncertainty in \MET is estimated by propagating the uncertainties in the transverse momenta of reconstructed objects and
by applying momentum scale and resolution uncertainties to the track-based soft term~\cite{Aaboud:2018tkc}.
A variation in the pile-up reweighting of MC events is included in order to cover the uncertainty in the ratio of the predicted and measured $pp$ inelastic cross sections~\cite{Aaboud:2016mmw}. 
For the measurements of the $W$ charge-dependent polarisation fractions, an uncertainty arising from the charge misidentification of electrons is also considered.

The uncertainty in the amount of background from misidentified leptons takes into account the limited number of events in the control regions as well as the difference in background composition between the control region used to determine the lepton misidentification rate and the control regions used to estimate the yield in the signal region.
This results in an uncertainty of about $25\%$ in the total  misidentified-lepton background yield and in the shape of the differential distributions of the reducible background events.
 
Uncertainties due to the theoretical modelling of \zz generated events are considered.
They arise from higher-order QCD corrections and the PDFs and are evaluated in the same way as for \wz\ events.
The uncertainty due to irreducible background sources other than \zz is evaluated by propagating the uncertainty in their MC cross sections.
These are $30\%$ for $VVV$~\cite{ATL-PHYS-PUB-2016-002}, $25$\% for \wzjjew~\cite{WZVBS36fb} and $15\%$ for $tZj$~\cite{Aad:2020wog} and $t\bar{t}+V$~\cite{ATLAS:2021fzm,ATL-PHYS-PUB-2016-005}.
The contribution of events from \wz\ production with at least one $\tau$-lepton decay is linked in the fit to the measured \wz\ production cross section for  electron and muon decays.
A normalisation uncertainty of $10$\% is attributed to this contribution to account for a possible mis-modelling in the MC simulation of the efficiency to select \wz\ events with a $\tau$-lepton decay.

The uncertainty in the combined 2015--2018 integrated luminosity is 1.7\%~\cite{ATLAS-CONF-2019-021}, obtained using the LUCID-2 detector~\cite{LUCID2} for the primary luminosity measurements.
It is applied to the signal normalisation as well as to all background contributions that are estimated
using only MC simulations.

The effect of the systematic uncertainties on the final results after the maximum-likelihood fit is shown in Table~\ref{tab:2VResults:SystematicSummary_pol} where the breakdown of the contributions to the uncertainties in the measured joint polarisation fractions is presented.


\section{Results}

\subsection{Joint boson polarisation measurements}

\begin{figure}[!htbp]
\begin{center}
\includegraphics[width=.8\textwidth]{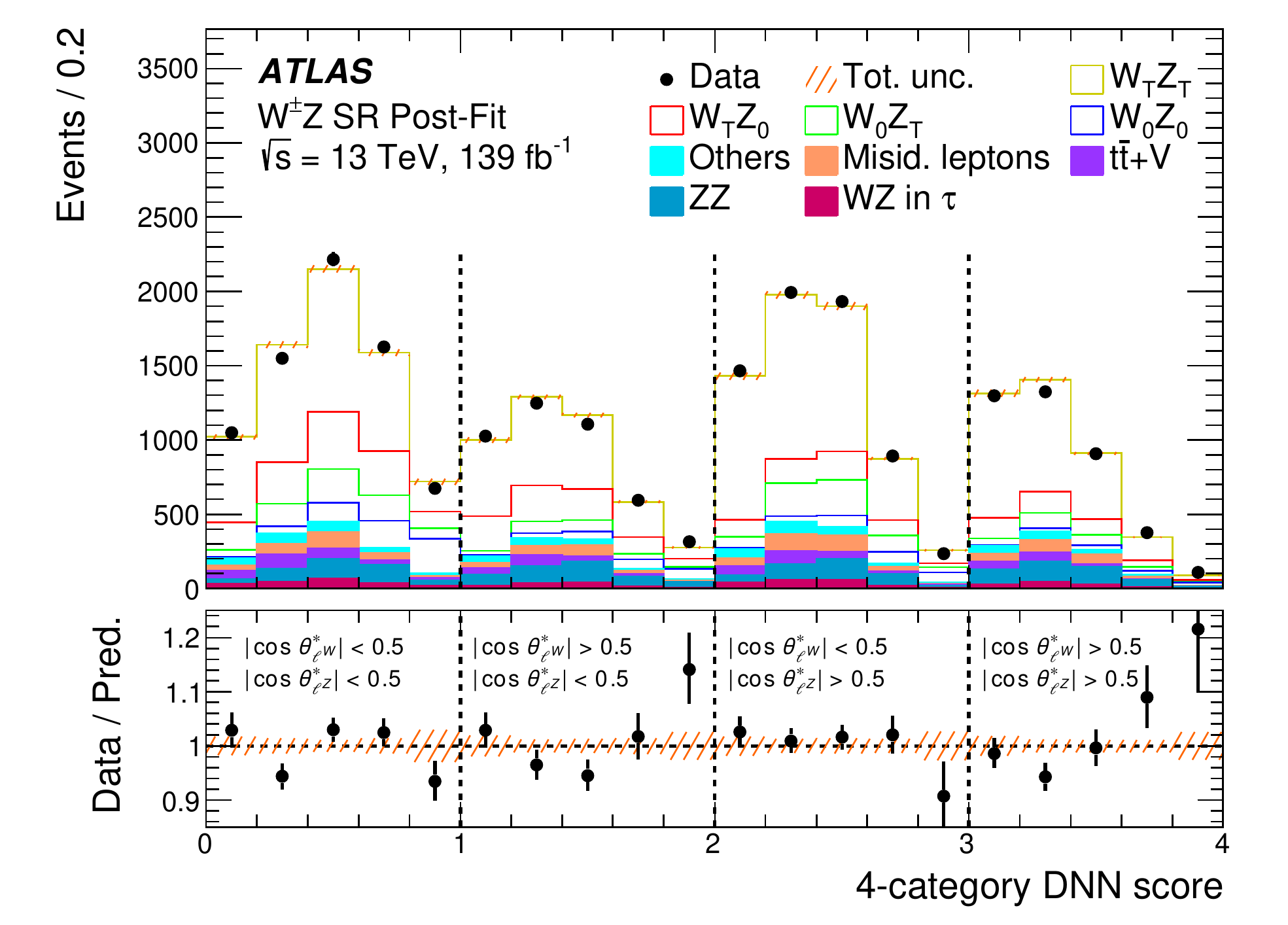}
\caption{
Post-fit distribution of the \PolCatDNNscore in \wz\ events at detector level in the signal region (SR).
The \PolCatDNNscore is defined for each event as the sum of the \PolDNNscore and of the category number ranging from 0 to 3.
Respective contributions of the \wz\ joint-helicity states 00, 0T, T0 and TT and backgrounds are normalised to the expected number of events after the fit.
The sum of background events containing misidentified leptons is labelled `Misid. leptons'.
The sum of $VVV$, \wzjjew and \tz background events is labelled `Others'.
The uncertainty band around the expectation includes all systematic uncertainties as obtained from the fit.
}
\label{fig:2VResults:postFitDist}
\end{center}
\end{figure}

The distribution of the \PolCatDNNscore in the \wz\ signal region  with background normalisations, signal normalisation and nuisance parameters adjusted by the profile-likelihood fit is shown in Figure~\ref{fig:2VResults:postFitDist}.
The corresponding pre-fit and post-fit yields in the signal region and in the $ZZ$ control region are detailed in Table~\ref{tab:2VResults:YieldTable}.

\begin{table}[!htbp]
\caption{
Expected and observed numbers of events
in the  $W^{\pm}Z$ signal region (left) and in the $ZZ$ control region (right).
Numbers are presented before and after a fit to the \PolCatDNNscore distribution in $W^{\pm}Z$ events.
The sum of background events containing misidentified leptons is labelled `Misid. leptons'.
The sum of $VVV$, \wzjjew and \tz background events is labelled `Others'.
The pre-fit uncertainties quoted are of statistical and systematic
origin but do not take into account correlations between parameters of the
fit.
The post-fit uncertainties quoted are of statistical and systematic
origin and take into account the correlations between the various
parameters of the fit.
For the \wz\ signal templates, the correlations arising from the relationship $\fzz + \fzT + \fTz + \fTT = 1$ explain the larger post-fit uncertainties compared to the pre-fit ones.
}
\label{tab:2VResults:YieldTable}
\centering
\resizebox{0.99\textwidth}{!}{
\begin{tabular}{l r}
\begin{tabular}{l r@{\,}l!{$\pm$}r@{\,}l r@{\,}l!{$\pm$}r@{\,}l}
\toprule
& \multicolumn{8}{c}{Signal Region} \\
& \multicolumn{4}{c}{Pre-fit} & \multicolumn{4}{c}{Post-fit} \\
\midrule
$WZ$ in $\tau$ & 620& & 60& & 630& & 60& \\
$ZZ$ & 1420& & 120& & 1630& & 50& \\
$t\bar{t}+V$ & 870& & 130& & 820& & 120& \\
Misid. leptons & 1170& & 230& & 1010& & 220& \\
Others & 800& & 90& & 780& & 90& \\
$W_{0}Z_{0}$ & 920& & 40& & 1190& & 160& \\
$W_{0}Z_{\mathrm{T}}$ & 2670& & 50& & 1900& & 500& \\
$W_{\mathrm{T}}Z_{0}$ & 2670& & 60& & 3100& & 400& \\
$W_{\mathrm{T}}Z_{\mathrm{T}}$ & 10200& & 230& & 10900& & 600& \\
\midrule
Total MC & 21400& & 500& & 21950& & 170& \\
\midrule
Data & \multicolumn{4}{c}{---} & \multicolumn{4}{c}{21936} \\
\bottomrule
\end{tabular}
&
\begin{tabular}{l r@{\,}l!{$\pm$}r@{\,}l r@{\,}l!{$\pm$}r@{\,}l}
\toprule
& \multicolumn{8}{c}{$ZZ$ Control Region} \\
& \multicolumn{4}{c}{Pre-fit} & \multicolumn{4}{c}{Post-fit} \\
\midrule
$WZ$ unpol. & 35&.6 & 1&.9 & 35&.6 & 1&.9 \\
$ZZ$ & 2030& & 150& & 2290& & 50& \\
$t\bar{t}+V$ & 153& & 23& & 143& & 21& \\
Misid. leptons & 14& & 4& & 15& & 4& \\
Others & 32& & 8& & 33& & 8& \\
\midrule
Total MC & 2260& & 150& & 2510& & 50& \\
\midrule
Data & \multicolumn{4}{c}{---} & \multicolumn{4}{c}{2554} \\
\bottomrule
\end{tabular}
\end{tabular}
}
\end{table}

The measurements of \fzz, \fzT, \fTz and \fTT are summarised in Table~\ref{tab:2VResults:PolRes}.
The hypotheses of having no events in each of the joint helicity states are tested.
In \wz events, the presence of a pair of $W$ and $Z$ bosons with a simultaneous longitudinal polarisation (\fzz) is observed  with a significance of $7.1 \,\sigma$, compared to $6.2 \, \sigma$ expected.
The other joint helicity fractions \fzT, \fTz and \fTT are also measured with observed (expected) significances of $3.4 \,\sigma$ ($5.4 \,\sigma$), $7.1 \,\sigma$ ($6.6 \,\sigma$) and $11 \,\sigma$ ($9.7 \,\sigma$), respectively.
The joint helicity fractions are also measured in \wpz and \wmz events separately, with observed significances for \fzz of  $6.9 \, \sigma$ and $4.1 \, \sigma$, respectively.
Table~\ref{tab:2VResults:SystematicSummary_pol} shows the main sources of uncertainty in the measurement of the joint helicity fractions.
Uncertainties related to data statistics and to QCD effects, namely the QCD scale and modelling uncertainties, make similar contributions to the precision of the measurements.
The modelling uncertainty is dominated by the uncertainties in the shape of the distributions of polarised \wz\ events due to NLO QCD corrections.
Among them, the uncertainty from the DNN reweighting method is dominant, except for \fzz where the uncertainty from the template generation method is of equal size.
Different correlation schemes between bins and between helicity states are tested for this latter modelling uncertainty.
All yield consistent results.
Uncertainties in the energy scale and reconstruction of jets are dominant among detector uncertainties, showing how observables for joint polarisation are linked to jet emissions.
The values of \fzz, \fzT, \fTz  and \fTT measured in \wz\ events are shown in Figure~\ref{fig:2VResults:pol_Wpm}.
The global level of agreement between the four measured joint helicity fractions with the fixed-order predictions at NLO QCD~\cite{Denner:2020wz} is observed to be 1.4 $\sigma$.
A similar level of agreement is observed with respect to \powhegpythia predictions.
The impact of EW corrections on the joint helicity fractions is negligible, smaller than $3$\%~\cite{Le:2022lrp}.

\begin{table}[!htbp]
\caption{
Measured joint helicity fractions \fzz, \fzT, \fTz and \fTT
in the fiducial phase space, for \wz, $W^+Z$ and $W^-Z$ events.
The total uncertainties in the measurements are reported.
The measurements are compared with predictions from \powhegpythia and
from \NLO QCD fixed-order calculations~\cite{Denner:2020wz}.
The uncertainties in the \powhegpythia prediction include statistical, PDF
and QCD scale uncertainties; the uncertainties in the \NLO QCD fixed-order
prediction include QCD scale uncertainties.
}
\label{tab:2VResults:PolRes}
\centering

 
\begin{tabular}{l
r@{\,}l!{$\pm$}r@{\,}l r@{\,}l!{$\pm$}r@{\,}l r@{\,}l!{$\pm$}r@{\,}l
}
\toprule
 
& \multicolumn{4}{c}{Data} & \multicolumn{4}{c}{\textsc{Powheg+Pythia}} & \multicolumn{4}{c}{\textsc{NLO QCD}}\\
 
\midrule
 
\multicolumn{13}{c}{$W^{\pm}Z$} \\
 
\midrule
 
$f_{00}$ &
0&.067 & 0&.010&
0&.0590 & 0&.0009&
0.&058 & 0.&002
\\
 
$f_{0\mathrm{T}}$ &
0&.110 & 0&.029&
0&.1515 & 0&.0017&
0.&159 & 0.&003
\\
 
$f_{\mathrm{T}0}$ &
0&.179 & 0&.023&
0&.1465 & 0&.0017&
0.&149 & 0.&003
\\
 
$f_{\mathrm{TT}}$ &
0&.644 & 0&.032&
0&.6431 & 0&.0021&
0.&628 & 0.&004
\\
\midrule
 
\multicolumn{13}{c}{$W^{+}Z$} \\
 
\midrule
$f_{00}$ &
0&.072 & 0&.016&
0&.0583 & 0&.0012&
0.&057 & 0.&002
\\
$f_{0\mathrm{T}}$ &
0&.119 & 0&.034&
0&.1484 & 0&.0022&
0.&155 & 0.&003
\\
$f_{\mathrm{T}0}$ &
0&.152 & 0&.033&
0&.1461 & 0&.0022&
0.&147 & 0.&003
\\
$f_{\mathrm{TT}}$ &
0&.66 & 0&.04&
0&.6472 & 0&.0026&
0.&635 & 0.&004
\\
\midrule
 
\multicolumn{13}{c}{$W^{-}Z$} \\
 
\midrule
$f_{00}$ &
0&.063 & 0&.016&
0&.0600 & 0&.0014&
0.&059 & 0.&002
\\
$f_{0\mathrm{T}}$ &
0&.11 & 0&.04&
0&.1560 & 0&.0027&
0.&166 & 0.&003
\\
$f_{\mathrm{T}0}$ &
0&.21 & 0&.04&
0&.1470 & 0&.0027&
0.&152 & 0.&003
\\
$f_{\mathrm{TT}}$ &
0&.62 & 0&.05&
0&.6370 & 0&.0033&
0.&618 & 0.&004
\\
 
\bottomrule
\end{tabular}


\end{table}

\begin{table}[!htbp]
\caption{
Summary of the relative uncertainties in the joint helicity fractions
\fzz, \fzT, \fTz, and \fTT measured in \wz\ events. The uncertainties are reported as percentages.
}
\label{tab:2VResults:SystematicSummary_pol}
\begin{center}

\begin{tabular}{l r@{\,}l r@{\,}l r@{\,}l r@{\,}l}
\toprule
&  \multicolumn{2}{c}{\fzz}  & \multicolumn{2}{c}{\fzT} & \multicolumn{2}{c}{\fTz} & \multicolumn{2}{c}{\fTT} \\
\midrule
\multicolumn{9}{c}{ Relative uncertainty [\%] }\\
\midrule
$e$ energy scale and id. efficiency        & 0&.34     & 0&.6      & 0&.8      & 0&.31     \\
$\mu$ energy scale and id. efficiency      & 0&.8      & 0&.23     & 0&.23     & 0&.13     \\
$E_{\mathrm{T}}^{\mathrm{miss}}$ and jets  & 3&.3      & 1&.3      & 1&.2      & 0&.4      \\
Pile-up                                    & 0&.6      & 0&.17     & 0&.4      & 0&.15     \\
Misidentified lepton background            & 2&.3      & 1&.6      & 0&.8      & 0&.26     \\
$ZZ$ background                            & 0&.9      & 0&.17     & 0&.32     & 0&.07     \\
Other backgrounds                          & 3&.0      & 1&.6      & 1&.3      & 0&.4      \\
\midrule
Parton Distribution Function               & 0&.5      & 1&.8      & 0&.09     & 0&.5      \\
QCD scale                                  & 0&.19     & 8&        & 0&.9      & 2&.0      \\
Modelling                                  & 9&        & 4&        & 2&.9      & 1&.2      \\
\midrule
Total systematic uncertainty               & 14&       & 15&       & 8&        & 4&        \\
Luminosity                                 & 0&.35     & 0&.24     & 0&.15     & 0&.05     \\
Statistical uncertainty                    & 13&       & 10&       & 12&       & 3&.0      \\
\midrule
Total                                      & 19&       & 18&       & 14&       & 5&        \\
 
\bottomrule
\end{tabular}


\end{center}
\end{table}

\begin{figure}[!htbp]
\centering
\includegraphics[width=.70\textwidth]{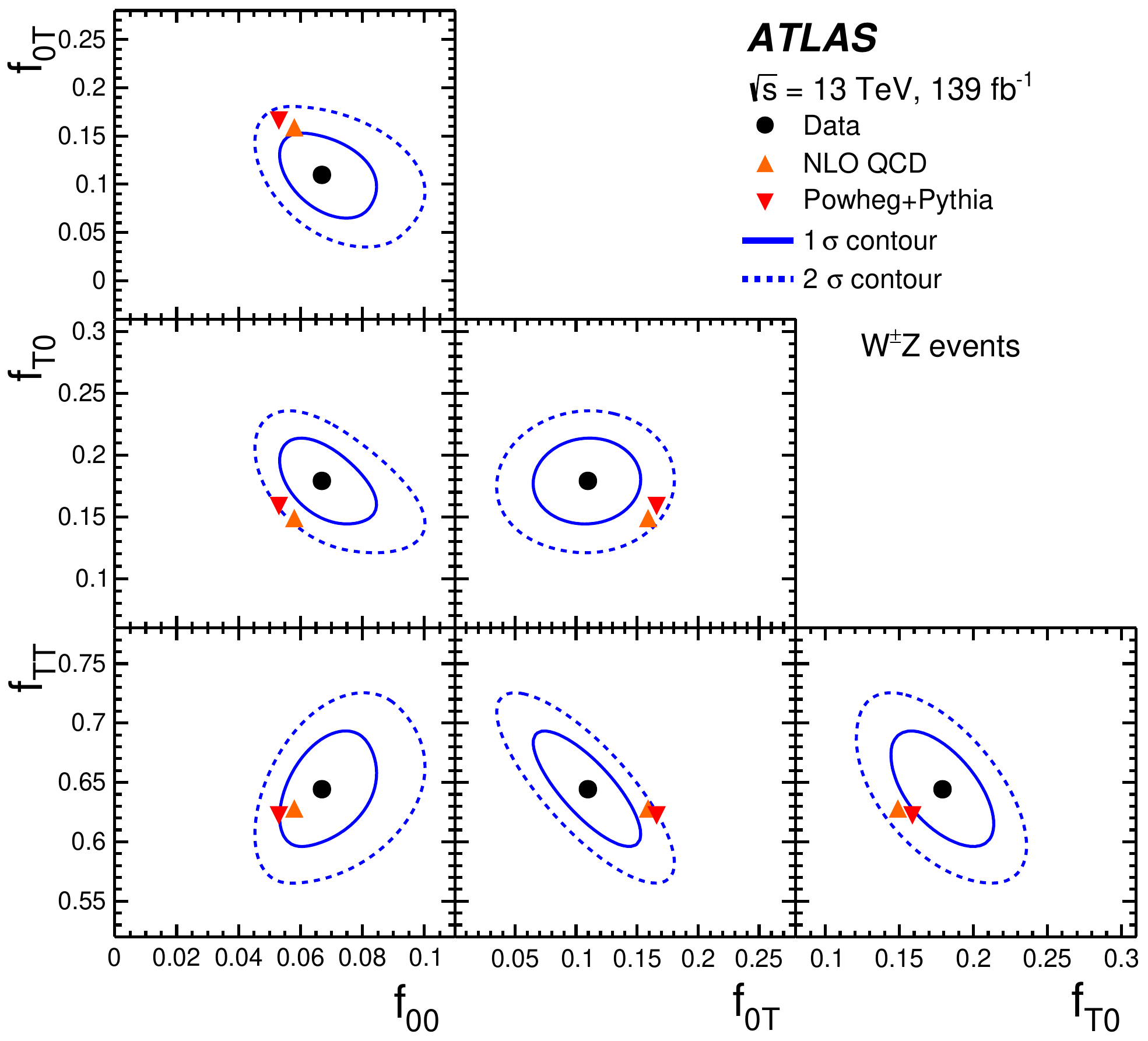}
\begin{center}
\caption{
Measured joint helicity fractions \fzz, \fzT, \fTz and \fTT of the $W$ and $Z$
bosons in \wz events,
compared with NLO QCD fixed-order predictions~\cite{Denner:2020wz} (upward pointing triangle) and to MC predictions from \powhegpythia (downward pointing triangle).
The effect of PDF and QCD scale uncertainties on the \powhegpythia and
NLO QCD fixed-order predictions are of the same size as the respective markers.
The full and dashed ellipses around the data points correspond to one
and two standard deviations, respectively.
}
\label{fig:2VResults:pol_Wpm}
\end{center}
\end{figure}


\subsection{Individual boson polarisation measurements}

 
The distributions of  \qctw and \ctz  with background normalisations, signal normalisation and nuisance parameters, adjusted by the profile-likelihood fit, are shown in Figure~\ref{fig:1VResults:postFitDist}.
The post-fit event yields in the \wz signal region and in the \zz control region are compatible with those obtained in the fit to the \PolCatDNNscore distribution.

\begin{figure}[!htbp]
\begin{center}
\includegraphics[width=.45\textwidth]{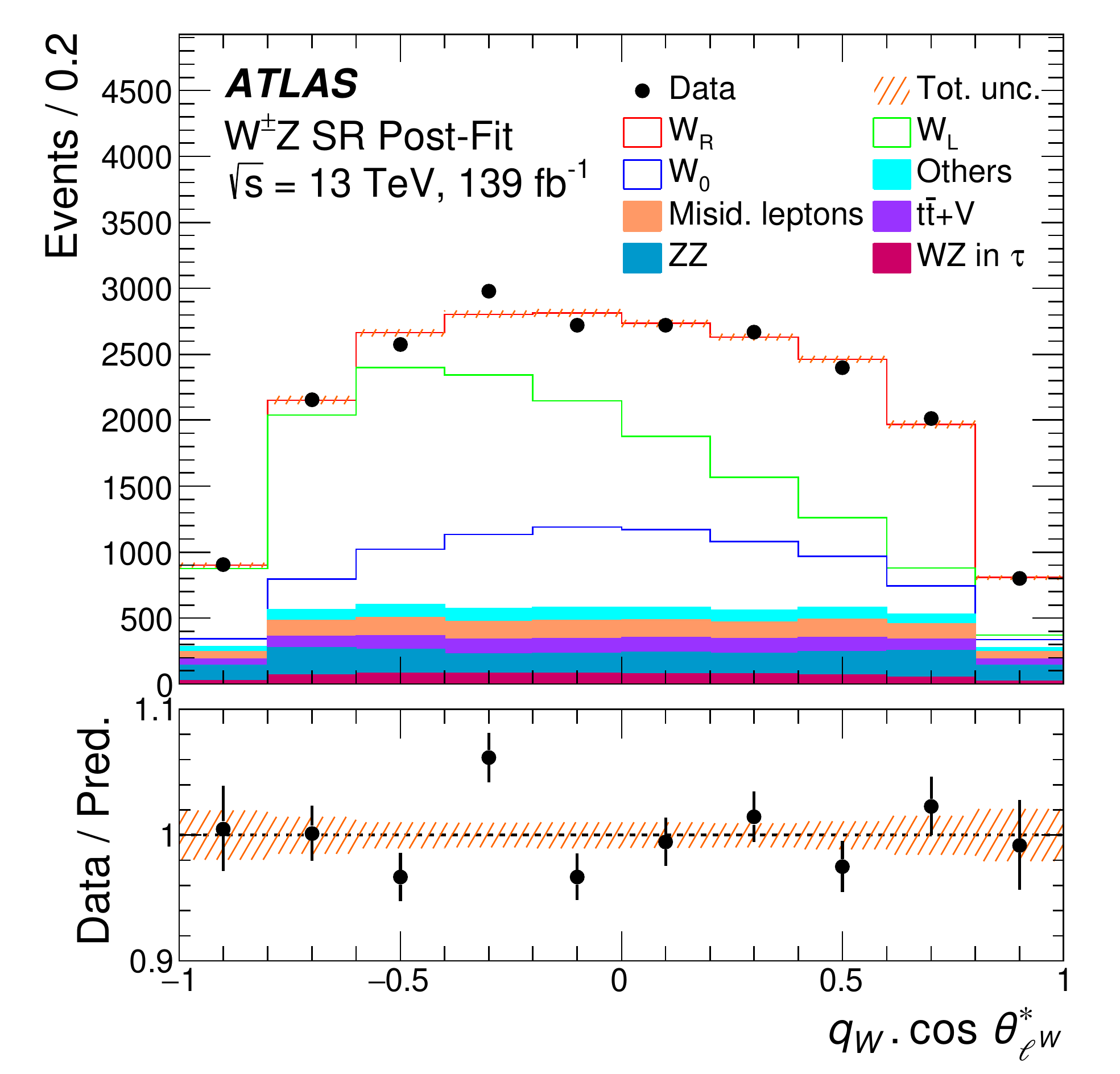}\put(-150,150){{(a)}}
\includegraphics[width=.45\textwidth]{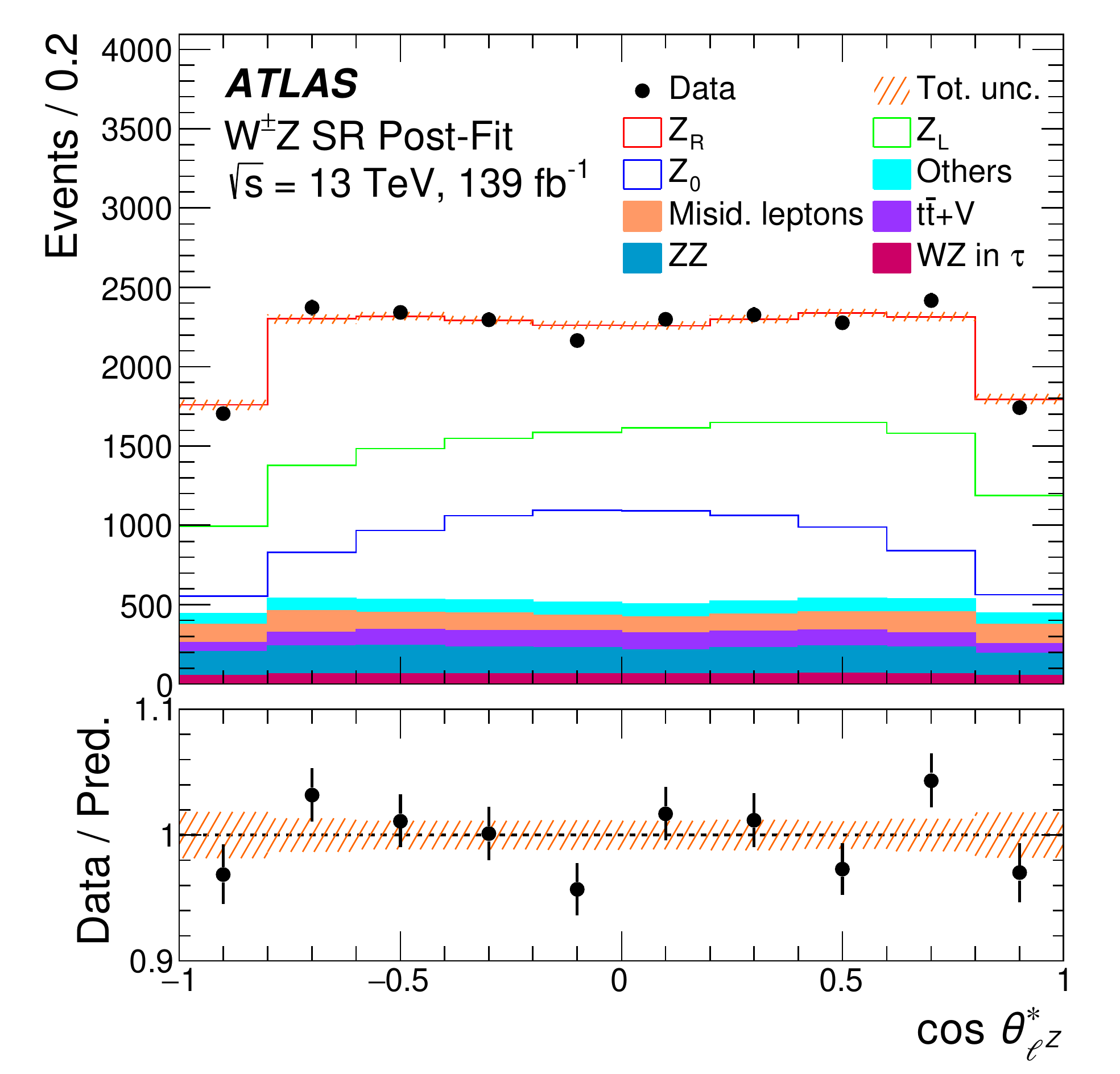}\put(-150,150){{(b)}}
\caption{
Post-fit distributions of (a) \qctw and (b) \ctz in $W^{\pm}Z$ events at
detector level in the signal region (SR).
Respective contributions of the \wz\ individual helicity states 0, L and R of the $W$ and $Z$ bosons and backgrounds are normalised to the expected number of events after the fit.
The sum of background events containing misidentified leptons is labelled `Misid. leptons'.
The sum of $VVV$, \wzjjew and \tz background events is labelled `Others'.
The uncertainty band around the expectation includes all systematic uncertainties as obtained from the fit.
}
\label{fig:1VResults:postFitDist}
\end{center}
\end{figure}

The measurements of \f0 and \fLR are summarised in Table~\ref{tab:1VResults:PolRes} where they are compared with the predictions from \powhegpythia and, for \f0, from an NLO QCD fixed-order calculation~\cite{Denner:2020wz}.
Good agreement of the measured helicity fractions of both the $W$ and $Z$ bosons with the predictions from \powhegpythia is observed.
Measured \f0 values agree within $1 \sigma$ with the prediction, while \fLR values agree within $1.5 \, \sigma$, except for $W^-Z$ events where the level of agreement deteriorates to $2.3 \, \sigma$.
Table~\ref{tab:1VResults:SystematicSummary_pol} shows the main sources of uncertainty in the measurement of the individual helicity fractions.
The measurements are dominated by statistical uncertainties, except for the \f0 measurement of the $W$ boson where the modelling uncertainty is equally important.
The values of \f0 and \fLR measured in \wz\ events are shown in Figure~\ref{fig:1VResults:pol_Wpm} for the $W$ and $Z$ bosons.

\begin{table}[!htbp]
\caption{
Individual helicity fractions \f0 and \fLR of the $W$ and $Z$
bosons measured in the fiducial phase space for \wz, $W^+Z$ and $W^-Z$ events.
The total uncertainties in the measurements are reported.
The measurements are compared with predictions from \powhegpythia and, for
\f0, from NLO QCD fixed-order calculations~\cite{Denner:2020wz}.
The uncertainties in the \powhegpythia prediction include statistical, PDF
and QCD scale uncertainties; the uncertainties in the \NLO QCD fixed-order
prediction include QCD scale uncertainties.
}
\label{tab:1VResults:PolRes}
\centering
\resizebox{0.97\textwidth}{!}{
 
\begin{tabular}{l | r@{\,}l!{$\pm$}r@{\,}l r@{\,}l!{$\pm$}r@{\,}l r@{\,}l!{$\pm$}r@{\,}l |  r@{\,}l!{$\pm$}r@{\,}l r@{\,}l!{$\pm$}r@{\,}l }
\toprule
& \multicolumn{12}{c|}{\f0} & \multicolumn{8}{c}{\fLmR} \\
& \multicolumn{4}{c}{Data} & \multicolumn{4}{c}{\textsc{Powheg+Pythia}} & \multicolumn{4}{c|}{\textsc{NLO QCD}}
& \multicolumn{4}{c}{Data} & \multicolumn{4}{c}{\textsc{Powheg+Pythia}}  \\
\midrule
 
$W$ in $W^{+}Z$ &
0&.23 & 0&.05&
0&.2044 & 0&.0024&
0.&211 & 0.&002
&
0&.071 & 0&.023&
0&.0990 & 0&.0015\\
 
$W$ in $W^{-}Z$ &
0&.19 & 0&.05&
0&.217 & 0&.004&
0.&225 & 0.&001
&
0&.026 & 0&.027&
-0&.0491 & 0&.0020\\
 
$W$ in $W^{\pm}Z$ &
0&.21 & 0&.04&
0&.2094 & 0&.0016&
0.&217 & 0.&001
&
0&.059 & 0&.016&
0&.0390 & 0&.0011\\
 
$Z$ in $W^{+}Z$ &
0&.223 & 0&.025&
0&.1971 & 0&.0019&
0.&206 & 0.&002
&
-0&.20 & 0&.10&
-0&.217 & 0&.006\\
 
$Z$ in $W^{-}Z$ &
0&.241 & 0&.029&
0&.2065 & 0&.0023&
0.&211 & 0.&001
&
0&.10 & 0&.13&
0&.092 & 0&.007\\
 
$Z$ in $W^{\pm}Z$ &
0&.231 & 0&.019&
0&.2009 & 0&.0014&
0.&208 & 0.&001
&
-0&.10 & 0&.08&
-0&.092 & 0&.005\\
 
\bottomrule
\end{tabular}

}
\end{table}

\begin{table}[!htbp]
\caption{
Summary of the relative uncertainties in the polarisation fractions \f0
and \fLR measured in $W^{\pm}Z$ events for $W$ and $Z$ bosons. The uncertainties are reported as percentages.
}
\label{tab:1VResults:SystematicSummary_pol}
\centering
\footnotesize{

\begin{tabular}{l r@{\,}l r@{\,}l r@{\,}l r@{\,}l}
\toprule
& \multicolumn{4}{c}{$W^{\pm}$ in $W^{\pm}Z$} & \multicolumn{4}{c}{$Z$ in $W^{\pm}Z$} \\
&  \multicolumn{2}{c}{\f0}  & \multicolumn{2}{c}{\fLmR}
&  \multicolumn{2}{c}{\f0}  & \multicolumn{2}{c}{\fLmR} \\
\midrule
\multicolumn{9}{c}{Relative uncertainty [\%] } \\
\midrule
$e$ energy scale and id. efficiency        & 1&.4      & 0&.8      & 1&.3      & 0&.7      \\
$\mu$ energy scale and id. efficiency      & 2&.1      & 5&        & 0&.8      & 0&.5      \\
$E_{\mathrm{T}}^{\mathrm{miss}}$ and jets  & 1&.9      & 2&.8      & 0&.28     & 3&.0      \\
Pile-up                                    & 1&.4      & 4&        & 1&.2      & 3&.1      \\
Misidentified lepton background            & 3&.4      & 0&.8      & 1&.6      & 1&.2      \\
$ZZ$ background                            & 0&.7      & 0&.6      & 0&.6      & 2&.5      \\
Other backgrounds                          & 0&.9      & 1&.3      & 0&.7      & 1&.3      \\
\midrule
Parton Distribution Function               & 0&.5      & 2&.9      & 0&.05     & 0&.5      \\
QCD scale                                  & 6&        & 6&        & 0&.22     & 5&        \\
Modelling                                  & 12&       & 3&.1      & 2&.2      & 19&       \\
\midrule
Total systematic uncertainty               & 14&       & 11&       & 3&.5      & 21&       \\
Luminosity                                 & 0&.25     & 0&.09     & 0&.06     & 0&.19     \\
Statistical uncertainty                    & 13&       & 40&       & 9&        & 90&       \\
\midrule
Total                                      & 19&       & 40&       & 10&       & 90&       \\
 
\bottomrule
\end{tabular}


}
\end{table}

\begin{figure}[!htbp]
\begin{center}
\includegraphics[width=.45\textwidth]{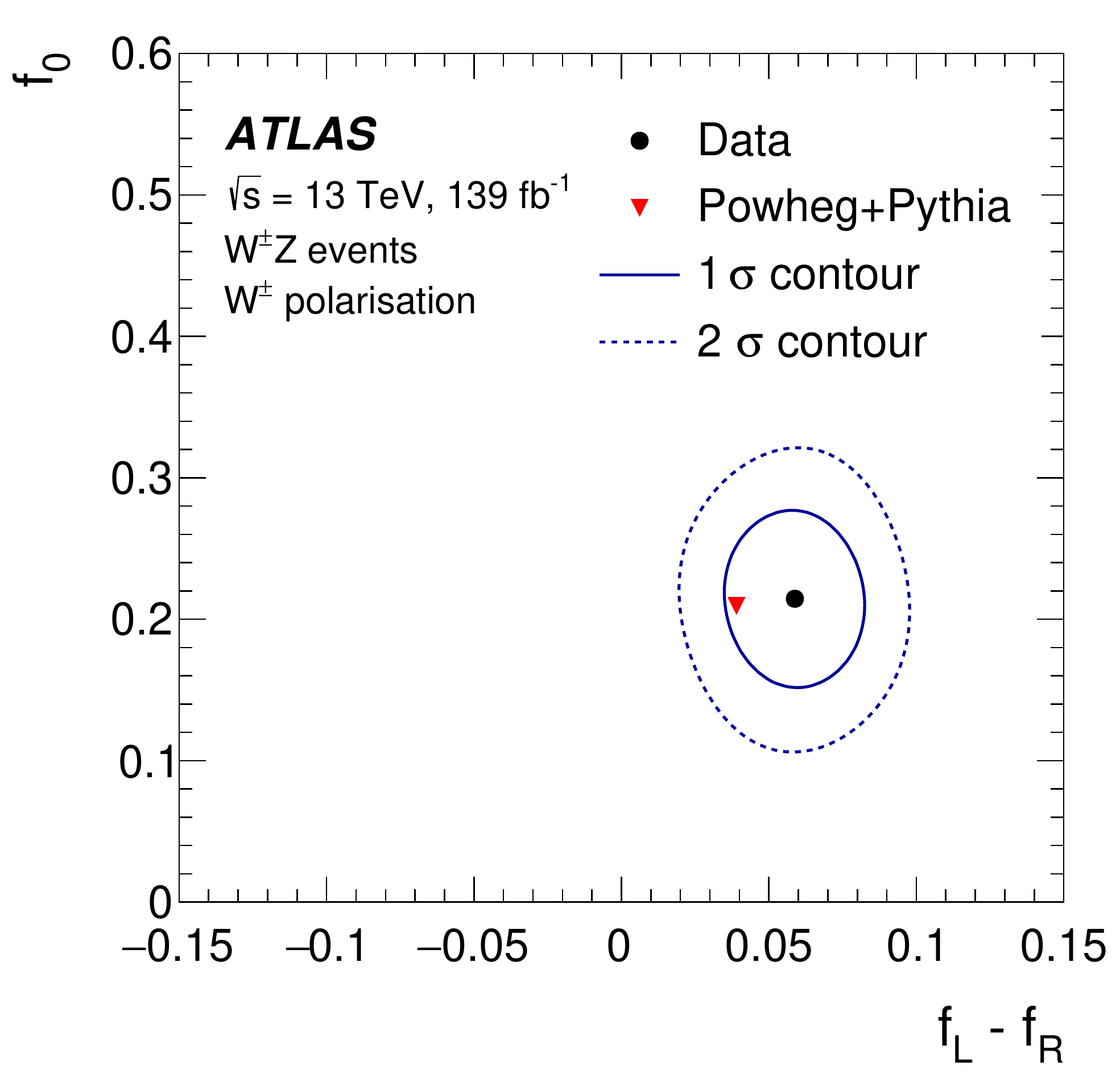}\put(-30,70){{(a)}}
\includegraphics[width=.45\textwidth]{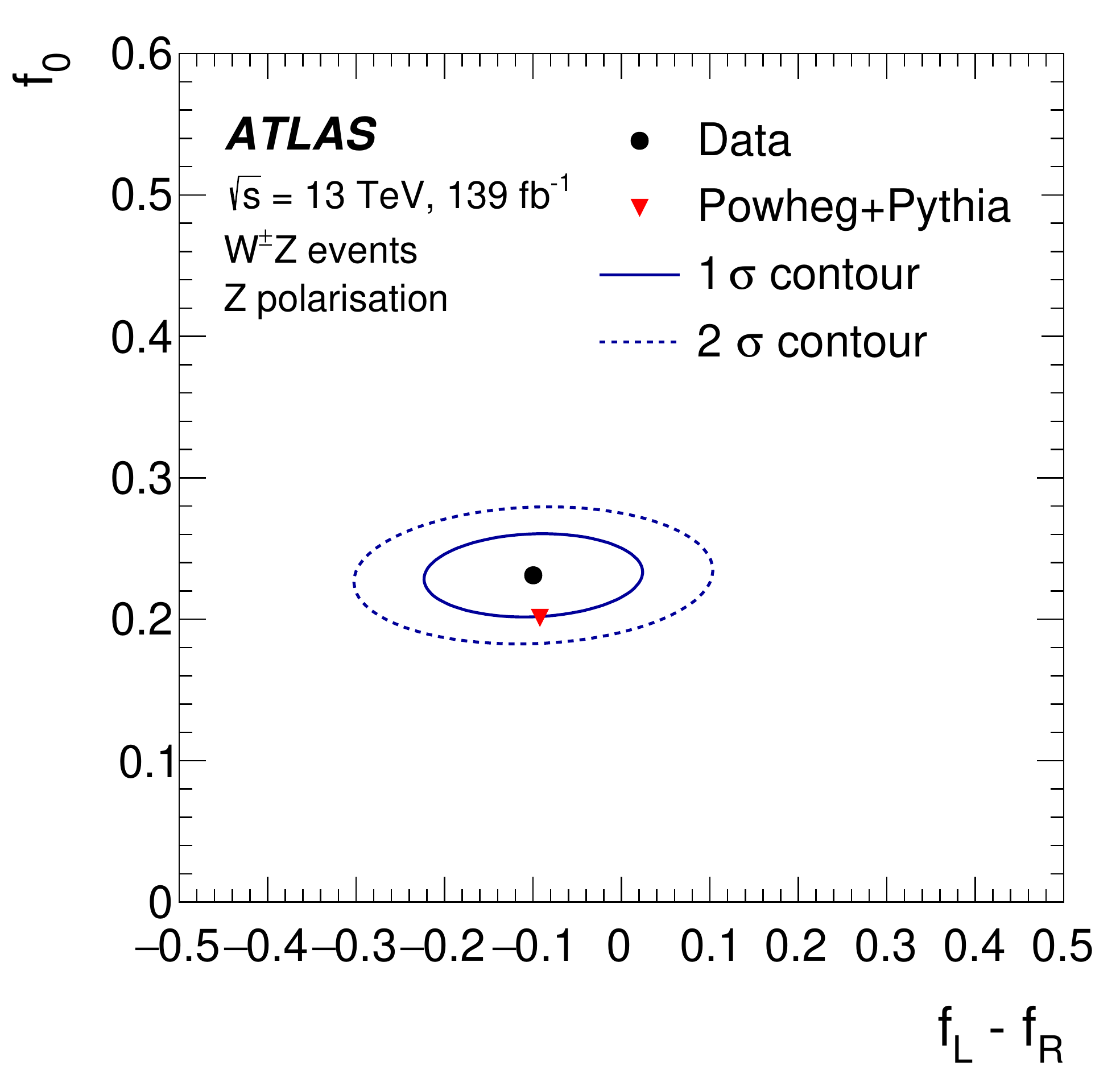}\put(-30,70){{(b)}}
\caption{
Individual helicity fractions   \f0 and \fLR measured for the (a) $W$ and (b)  $Z$ bosons in $W^{\pm}Z$ events, compared with predictions   from \powhegpythia with $\sin^2{\theta_{\mathrm{W}}}=0.23152$ (downward pointing triangle). 
The effect of PDF and QCD scale uncertainties on the \powhegpythia prediction 
are of the same size as the triangle marker.
The full and dashed ellipses around the data points correspond to one and two standard deviations, respectively.
}
\label{fig:1VResults:pol_Wpm}
\end{center}
\end{figure}

The compatibility of the measured joint and individual helicity fractions is tested.
Ignoring the interference between polarisations, which should represent only $0.6$\%--$0.8$\% of the events according to NLO QCD and NLO QCD+EW fixed-order predictions~\cite{Denner:2020wz,Le:2022lrp}, the following relationships ensue
\begin{equation}
\fzT = \f0^W - \fzz\, , \quad \fTz = \f0^Z - \fzz \quad \mathrm{and} \quad \fTT = 1 + \fzz - \f0^W - \f0^Z. \nonumber
\end{equation}
Using these relationships, joint \fzz and individual $\f0^W$, $\f0^Z$ helicity fractions can be measured simultaneously from a fit to the \PolCatDNNscore observable.
The fractions $\f0^W$ and $\f0^Z$ measured in this procedure agree within less than $1 \sigma$ of their own uncertainties alone  with the $\f0^W$ and $\f0^Z$ values obtained via individual \ctw and \ctz distributions.
This represents a consistency test of the joint helicity measurement procedure.
 
The spin-correlation between longitudinal $W$ and $Z$ bosons is tested by comparing joint and individual helicity fractions and measuring the  ratio $R_c =  \fzz / (\f0^W \f0^Z)$.
In the absence of spin correlations, neglecting interference effects, $R_c$ is equal to 1.
A value of $R_c = 1.3$ is predicted by NLO QCD fixed-order calculations, indicating the presence of correlation between polarisation states of the bosons.
The ratio $R_c$ is extracted from data by a simultaneous fit to the  \PolCatDNNscore observable of \fzz, $\f0^W$ and $\f0^Z$.
It is measured in \wz events to be $R_c = 1.54 \pm 0.35$.
The observed significance relative to the no spin-correlations hypothesis, i.e. $R_c = 1$, is $1.6 \, \sigma$.


\subsection{Differential cross-section measurements}

The inclusive cross section of \wz\ production in the fiducial region for leptonic decay modes (electrons or muons) is determined simultaneously to the joint helicity fractions.
It is measured to be
$\sigma_{W^\pm Z \rightarrow \ell^{'} \nu \ell \ell}^{\textrm{fid.}} =  64.6  \, \pm~0.5~\textrm{(stat.)} \, \pm~1.8~\textrm{(syst.)} \, \pm~1.1~\textrm{(lumi.)}$~fb,
where the uncertainties correspond to statistical uncertainties, experimental and modelling systematic uncertainties, and luminosity uncertainties.
The corresponding SM NNLO QCD prediction from \MATRIX~\cite{Grazzini:2017ckn}
is $64.0^{+1.5}_{-1.3} \, \mathrm{fb}$, where the uncertainty corresponds to the QCD scale uncertainty.

\begin{figure}[!htbp]
\begin{center}
\includegraphics[width=.42\textwidth]{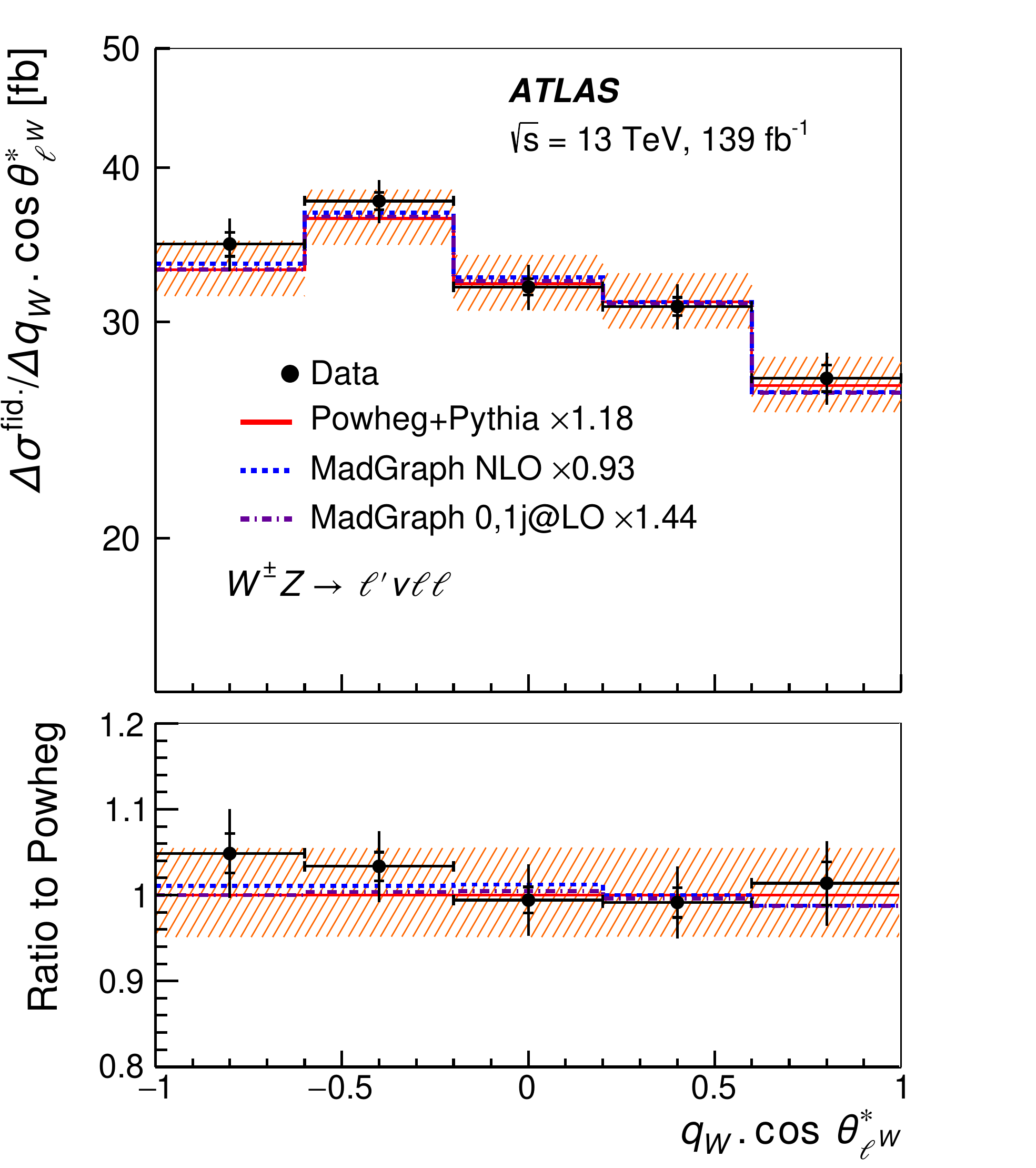}\put(-160,195){{(a)}}\quad
\includegraphics[width=.42\textwidth]{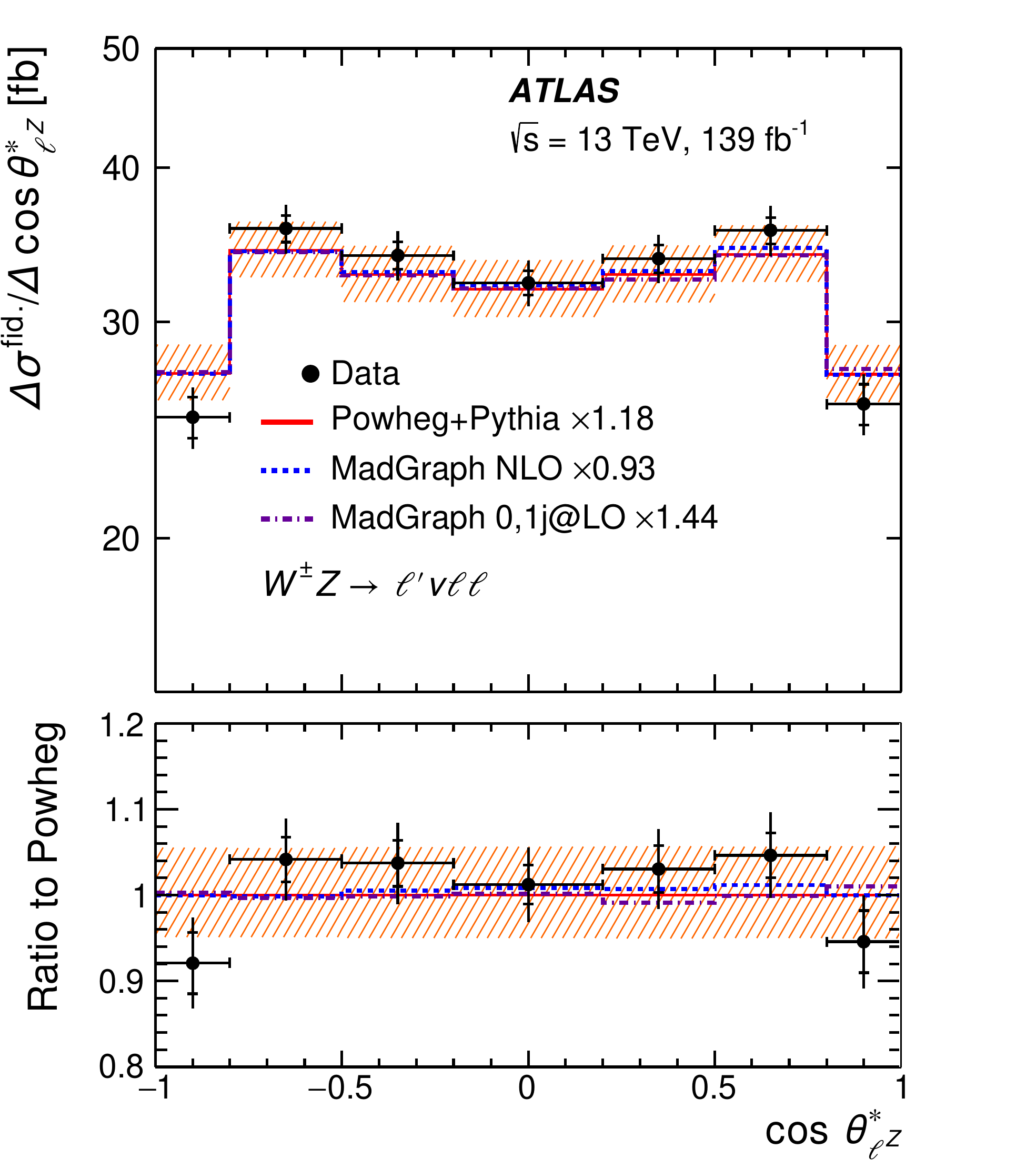}\put(-160,195){{(b)}}\\
\includegraphics[width=0.42\textwidth]{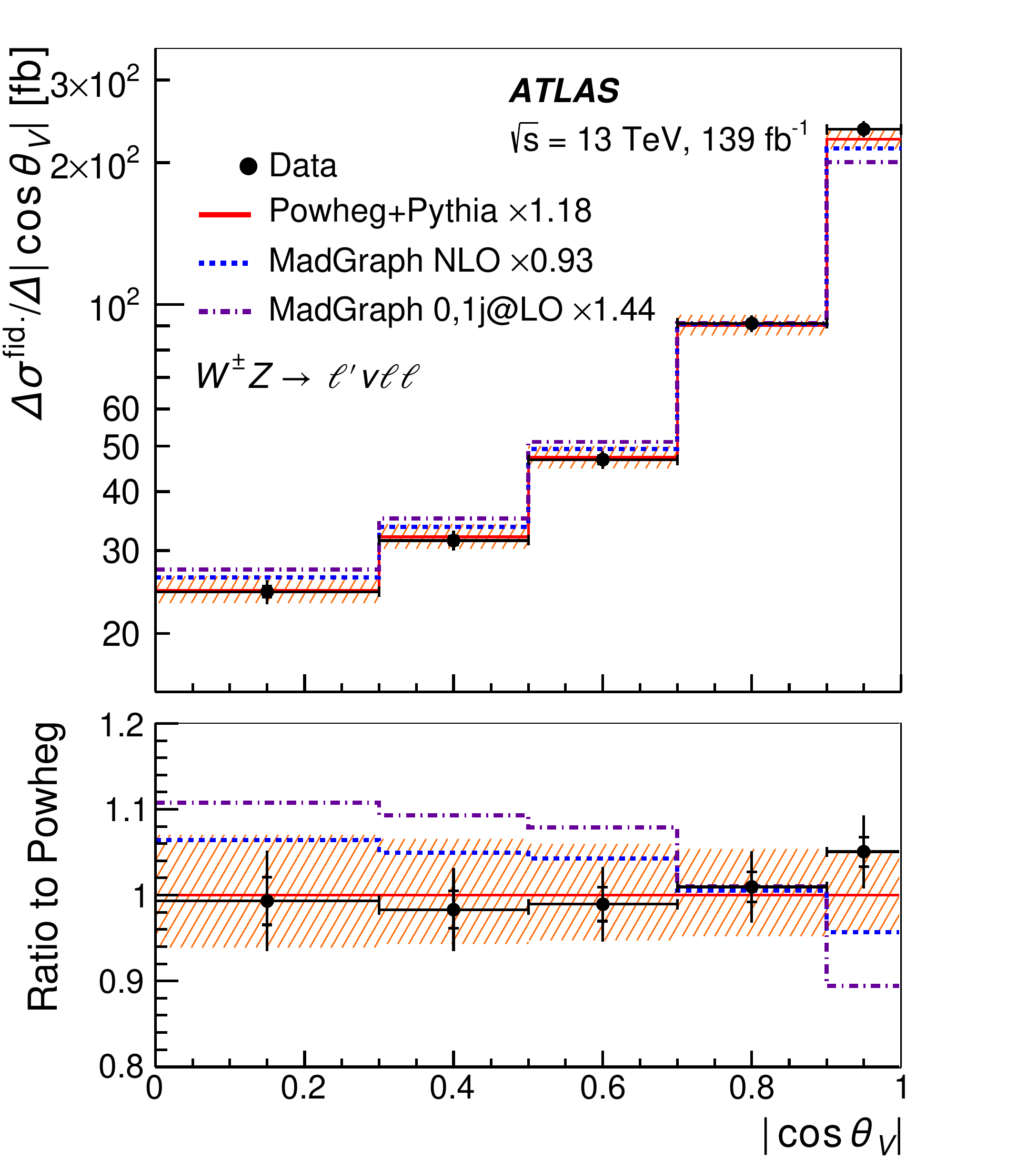}\put(-160,195){{(c)}}\quad
\includegraphics[width=0.42\textwidth]{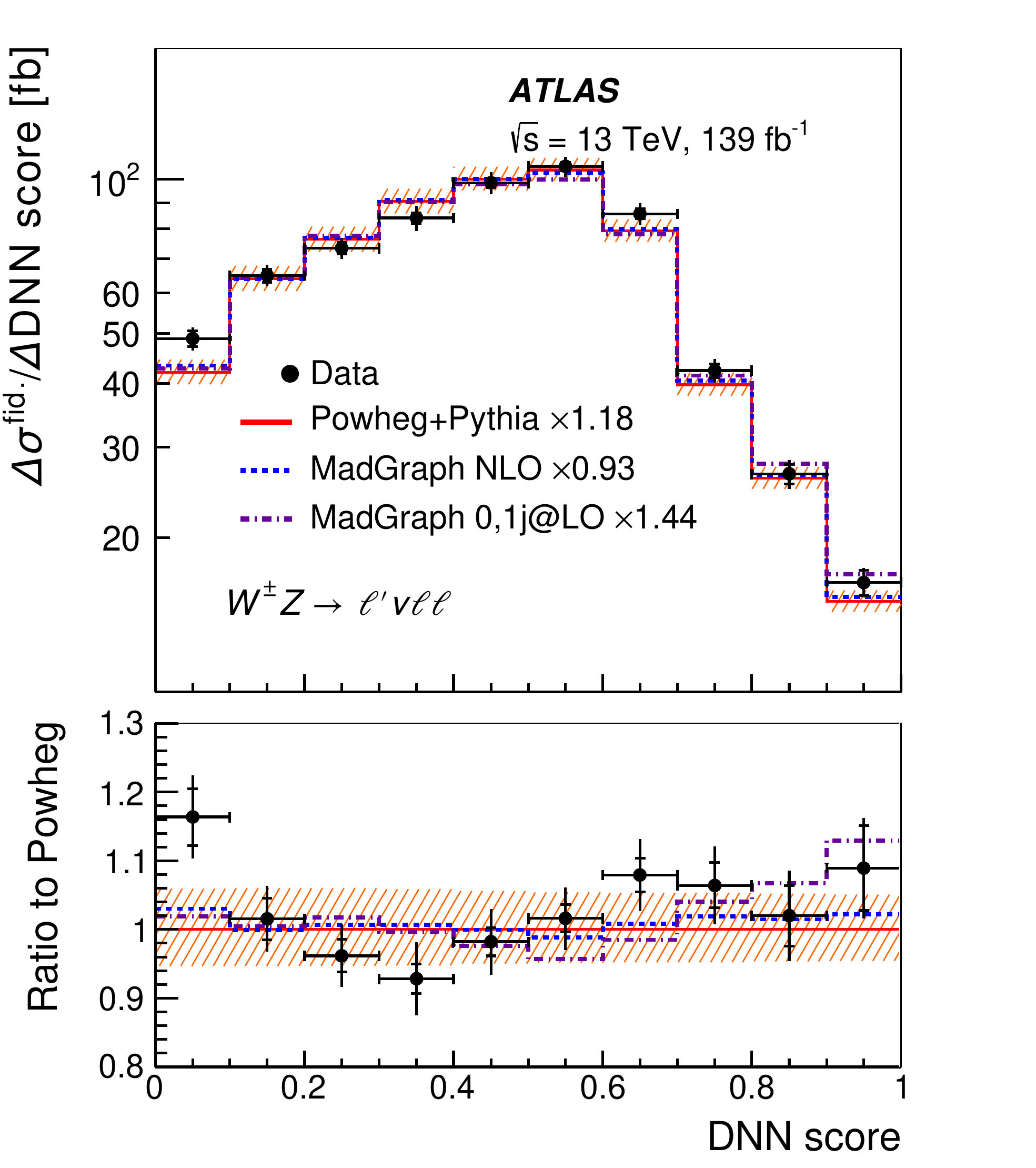}\put(-160,195){{(d)}}
\caption{The measured $W^{\pm}Z$ differential cross section in the fiducial phase space as a function of  (a) \qctw,  (b) \ctz, (c) \ctv  and (d) \PolDNNscore.
The total cross section integrated over each bin and divided by the bin width is represented.
The inner and outer error bars on the data points represent the statistical and total uncertainties, respectively.
The measurements are compared with the NLO prediction from \powhegpythia (solid line).
The dashed band shows how the sum of QCD scale and PDF uncertainties affects the \powhegpythia NLO prediction.
The predictions from the \mg01j and \Madgraph  MC generators are also indicated by dotted-dashed  and dashed  lines, respectively.
All predictions have been scaled to the NNLO QCD integrated cross section predicted by MATRIX.
}
\label{fig:DiffResults:diffDists}
\end{center}
\end{figure}

The \wz production cross section is  measured as a function of several variables sensitive to the individual and joint boson polarisation: \qctw, \ctz, \ctv and the \PolDNNscore.
The measured  differential cross sections in Figure~\ref{fig:DiffResults:diffDists} are compared with the Born-level predictions at NLO in QCD from the \powhegpythia, \Madgraph MC generators and to the sum of the polarised \MG01jLO predictions.
The Born-level \PolDNNscore prediction is evaluated using the same DNN as trained at detector level and evaluated using Born-level input observables.
The predicted integrated fiducial cross sections  of the three MC generators are rescaled to the NNLO cross section from \MATRIX~\cite{Grazzini:2017ckn}, as indicated in Figure~\ref{fig:DiffResults:diffDists}.
Good agreement of the shapes of the measured distributions with the NLO QCD predictions is observed.
The \MG01jLO prediction exhibits larger differences from the measured cross section for the \ctv distribution. 
This is expected as the \ctv distribution is particularly sensitive to NLO QCD corrections~\cite{Denner:2020wz} and the \MG01jLO prediction only includes their real parts.
The measured cross sections and the Rivet~\cite{Bierlich:2019rhm} routine for this measurement, including for the evaluation of the DNN classifier score, are published on HEPData~\cite{Maguire:2017ypu}\footnote{In the HEPData entry, two sets of cross section measurements are provided, determined using both dressed and Born leptons, respectively.}.


\FloatBarrier
 
\section{Conclusion}

Measurements of gauge boson polarisation states from \wz\ production  in $\sqrt{s} = 13$~\TeV\ $pp$ collisions at the LHC are presented.
The data analysed were collected with the ATLAS detector during the years $2015$--$2018$ and correspond to an integrated luminosity of $139~\ifb$.
The measurements use leptonic decay modes of the gauge bosons to electrons or muons and are performed in a fiducial phase space closely matching the detector acceptance.
 
Using neural-network-based advanced analysis techniques,  joint helicity fractions of pair-produced vector bosons are measured. 
For the first time, the presence of a $W$ boson and a $Z$ boson with a simultaneous longitudinal polarisation is measured with observed and expected significances of $7.1$ and $6.2$ standard deviations, respectively.
Integrated over the fiducial region, the fractions of the $W$ and $Z$ bosons both longitudinally or transversely polarised  are measured in \wz events to be $\fzz = 0.067 \pm 0.010$ and $\fTT = 0.644 \pm 0.032$, respectively, in agreement with the SM predictions at NLO in QCD.
The impact of EW corrections is negligible.
The measured fractions of mixed polarisation modes are  $\fzT = 0.110 \pm 0.029$ and $\fTz = 0.179 \pm 0.023$.
Joint helicity fractions are also measured separately for \wpz and \wmz events.

Individual helicity fractions \f0 and \fLmR of the $W$ and $Z$ bosons are measured. 
The measured values agree with the SM predictions and are consistent with the measured joint helicity fractions when neglecting interference among polarisation states,  which from NLO QCD+EW calculations are expected to be smaller than the uncertainty of the present measurement.

Furthermore, the inclusive cross section of \wz\ production in the fiducial region for leptonic decay modes is measured to be
$\sigma_{W^\pm Z \rightarrow \ell^{'} \nu \ell \ell}^{\textrm{fid.}} =  64.6  \, \pm~2.1$~fb,
in agreement with the NNLO SM expectation of $64.0^{+1.5}_{-1.3}$~fb.
The \wz\ production cross section is measured as a function of several kinematic variables sensitive to polarisation and compared with SM predictions at NLO  in QCD from the \powhegpythia and \Madgraph MC event generators.
In particular, the variable related to the output of the deep neural network classifier used for the joint helicity measurement is unfolded in order to remove detector effects.
The differential cross-section distributions are well described by the NLO MC predictions.


\section*{Acknowledgements}


We thank CERN for the very successful operation of the LHC, as well as the
support staff from our institutions without whom ATLAS could not be
operated efficiently.
 
We acknowledge the support of
ANPCyT, Argentina;
YerPhI, Armenia;
ARC, Australia;
BMWFW and FWF, Austria;
ANAS, Azerbaijan;
CNPq and FAPESP, Brazil;
NSERC, NRC and CFI, Canada;
CERN;
ANID, Chile;
CAS, MOST and NSFC, China;
Minciencias, Colombia;
MEYS CR, Czech Republic;
DNRF and DNSRC, Denmark;
IN2P3-CNRS and CEA-DRF/IRFU, France;
SRNSFG, Georgia;
BMBF, HGF and MPG, Germany;
GSRI, Greece;
RGC and Hong Kong SAR, China;
ISF and Benoziyo Center, Israel;
INFN, Italy;
MEXT and JSPS, Japan;
CNRST, Morocco;
NWO, Netherlands;
RCN, Norway;
MEiN, Poland;
FCT, Portugal;
MNE/IFA, Romania;
MESTD, Serbia;
MSSR, Slovakia;
ARRS and MIZ\v{S}, Slovenia;
DSI/NRF, South Africa;
MICINN, Spain;
SRC and Wallenberg Foundation, Sweden;
SERI, SNSF and Cantons of Bern and Geneva, Switzerland;
MOST, Taiwan;
TENMAK, T\"urkiye;
STFC, United Kingdom;
DOE and NSF, United States of America.
In addition, individual groups and members have received support from
BCKDF, CANARIE, Compute Canada and CRC, Canada;
PRIMUS 21/SCI/017 and UNCE SCI/013, Czech Republic;
COST, ERC, ERDF, Horizon 2020 and Marie Sk{\l}odowska-Curie Actions, European Union;
Investissements d'Avenir Labex, Investissements d'Avenir Idex and ANR, France;
DFG and AvH Foundation, Germany;
Herakleitos, Thales and Aristeia programmes co-financed by EU-ESF and the Greek NSRF, Greece;
BSF-NSF and MINERVA, Israel;
Norwegian Financial Mechanism 2014-2021, Norway;
NCN and NAWA, Poland;
La Caixa Banking Foundation, CERCA Programme Generalitat de Catalunya and PROMETEO and GenT Programmes Generalitat Valenciana, Spain;
G\"{o}ran Gustafssons Stiftelse, Sweden;
The Royal Society and Leverhulme Trust, United Kingdom.
 
The crucial computing support from all WLCG partners is acknowledged gratefully, in particular from CERN, the ATLAS Tier-1 facilities at TRIUMF (Canada), NDGF (Denmark, Norway, Sweden), CC-IN2P3 (France), KIT/GridKA (Germany), INFN-CNAF (Italy), NL-T1 (Netherlands), PIC (Spain), ASGC (Taiwan), RAL (UK) and BNL (USA), the Tier-2 facilities worldwide and large non-WLCG resource providers. Major contributors of computing resources are listed in Ref.~\cite{ATL-SOFT-PUB-2021-003}.


\printbibliography

@article{Gleisberg:2008fv,
      author         = "Gleisberg, Tanju and Hoeche, Stefan",
      title          = "{Comix, a new matrix element generator}",
      journal        = "JHEP",
      volume         = "12",
      year           = "2008",
      pages          = "039",
      doi            = "10.1088/1126-6708/2008/12/039",
      eprint         = "0808.3674",
      archivePrefix  = "arXiv",
      primaryClass   = "hep-ph",
      reportNumber   = "SLAC-PUB-13232, IPPP-08-31, DCPT-08-62, MCNET-08-08",
      SLACcitation   = "%%CITATION = ARXIV:0808.3674;%%"
}

@article{Schumann:2007mg,
      author         = "Schumann, Steffen and Krauss, Frank",
      title          = "{A Parton shower algorithm based on Catani-Seymour dipole
                        factorisation}",
      journal        = "JHEP",
      volume         = "03",
      year           = "2008",
      pages          = "038",
      doi            = "10.1088/1126-6708/2008/03/038",
      eprint         = "0709.1027",
      archivePrefix  = "arXiv",
      primaryClass   = "hep-ph",
      reportNumber   = "DCPT-07-86, IPPP-07-43",
      SLACcitation   = "%%CITATION = ARXIV:0709.1027;%%"
}

@article{Hoeche:2012yf,
      author         = "Hoeche, Stefan and Krauss, Frank and Schonherr, Marek and
                        Siegert, Frank",
      title          = "{QCD matrix elements + parton showers: The NLO case}",
      journal        = "JHEP",
      volume         = "04",
      year           = "2013",
      pages          = "027",
      doi            = "10.1007/JHEP04(2013)027",
      eprint         = "1207.5030",
      archivePrefix  = "arXiv",
      primaryClass   = "hep-ph",
      reportNumber   = "SLAC-PUB-15191, IPPP-12-52, DCPT-12-104, LPN12-081,
                        FR-PHENO-2012-017, MCNET-12-09, --FR-PHENO-2012-017",
      SLACcitation   = "%%CITATION = ARXIV:1207.5030;%%"
}

@article{Alwall:2014hca,
      author         = "Alwall, J. and Frederix, R. and Frixione, S. and Hirschi,
                        V. and Maltoni, F. and Mattelaer, O. and Shao, H. -S. and
                        Stelzer, T. and Torrielli, P. and Zaro, M.",
      title          = "{The automated computation of tree-level and
                        next-to-leading order differential cross sections, and
                        their matching to parton shower simulations}",
      journal        = "JHEP",
      volume         = "07",
      year           = "2014",
      pages          = "079",
      doi            = "10.1007/JHEP07(2014)079",
      eprint         = "1405.0301",
      archivePrefix  = "arXiv",
      primaryClass   = "hep-ph",
      reportNumber   = "CERN-PH-TH-2014-064, CP3-14-18, LPN14-066, MCNET-14-09,
                        ZU-TH-14-14",
      SLACcitation   = "%%CITATION = ARXIV:1405.0301;%%"
}

@Booklet{ATL-PHYS-PUB-2016-002,
    author         = "{ATLAS Collaboration}",
    title          = "{Multi-Boson Simulation for 13 TeV ATLAS Analyses}",
    howpublished   = "{ATL-PHYS-PUB-2016-002}",
    url            = "https://cds.cern.ch/record/2119986",
    year           = "2016",
}

@article{Denner:2020ww,
      author         = "Denner, Ansgar and Pelliccioli, Giovanni",
      title          = "{Polarized electroweak bosons in $W^+W^-$ production at the LHC including NLO
                         QCD effects}",
      journal        = "JHEP",
      volume         = "09",
      year           = "2020",
      pages          = "164",
      doi            = "10.1007/JHEP09(2020)164",
      eprint         = "2206.14867",
      archivePrefix  = "arXiv",
      primaryClass   = "hep-ph",
      reportNumber   = " ",
      SLACcitation   = "%%CITATION = ARXIV:1708.06938;%%"     
}

@article{Denner:2020wz,
      author         = "Denner, Ansgar and Pelliccioli, Giovanni",
      title          = "{NLO QCD predictions for doubly-polarized WZ production at the LHC}",
      journal        = "Phys. Lett. B",
      volume         = "814",
      year           = "2020",
      pages          = "136107",
      doi            = "10.1016/j.physletb.2021.136107",
      eprint         = "2010.07149",
      archivePrefix  = "arXiv",
      primaryClass   = "hep-ph",
      reportNumber   = " ",
      addendum   = "(We would like to thank Ansgar Denner and Giovanni Pellicioli for fruitful discussions and for providing us with the NLO QCD predictions for the specific deep-neural network classifier output used in the analysis and for both $W^+ Z$ and $W^- Z$ events.)",
      SLACcitation   = "%%CITATION = ARXIV:1708.06938;%%"                                                                                                   
}

@article{DAgostini:1994zf,
      author         = "D'Agostini, G.",
      title          = "{A multidimensional unfolding method based on Bayes'
                        theorem}",
      journal        = "Nucl. Instrum. Meth. A",
      volume         = "362",
      year           = "1995",
      pages          = "487",
      doi            = "10.1016/0168-9002(95)00274-X",
      reportNumber   = "DESY-94-099",
      SLACcitation   = "%%CITATION = NUIMA,A362,487;%%"
}

@article{RooUnfold,
      author         = "Adye, Tim",
      title          = "{Unfolding algorithms and tests using RooUnfold}",
      pages          = "313-318",
      year           = "2011",
      eprint         = "1105.1160",
      archivePrefix  = "arXiv",
      primaryClass   = "physics.data-an",
      SLACcitation   = "%%CITATION = ARXIV:1105.1160;%%",
}

@article{ATLAS:2012au,
      author         = "{ATLAS Collaboration}",
      title          = "{Measurement of the polarisation of $W$ bosons produced
                        with large transverse momentum in $pp$ collisions at
                        $\sqrt{s}=7$ TeV with the ATLAS experiment}",
      collaboration  = "ATLAS",
      journal        = "Eur. Phys. J. C",
      volume         = "72",
      year           = "2012",
      pages          = "2001",
      doi            = "10.1140/epjc/s10052-012-2001-6",
      eprint         = "1203.2165",
      archivePrefix  = "arXiv",
      primaryClass   = "hep-ex",
      reportNumber   = "CERN-PH-EP-2012-016",
      SLACcitation   = "%%CITATION = ARXIV:1203.2165;%%"
}

@techreport{TruthWS,
      author       = "{ATLAS Collaboration}",
      title         = "{Proposal for truth particle observable definitions in
                       physics measurements}",
      institution   = "CERN",
      address       = "Geneva",
      number        = "ATL-PHYS-PUB-2015-013",
      month         = "Jun",
      year          = "2015",
      reportNumber  = "ATL-PHYS-PUB-2015-013",
      url           = "https://cds.cern.ch/record/2022743",
}

@article{Butterworth:2015oua,
      author         = "Butterworth, Jon and others",
      title          = "{PDF4LHC recommendations for LHC Run II}",
      journal        = "J. Phys.",
      volume         = "G43",
      year           = "2016",
      pages          = "023001",
      doi            = "10.1088/0954-3899/43/2/023001",
      eprint         = "1510.03865",
      archivePrefix  = "arXiv",
      primaryClass   = "hep-ph",
      reportNumber   = "OUTP-15-17P, SMU-HEP-15-12, TIF-UNIMI-2015-14,
                        LCTS-2015-27, CERN-PH-TH-2015-249",
      SLACcitation   = "%%CITATION = ARXIV:1510.03865;%%"
}

@article{Azatov:2017kzw,
      author         = "Azatov, A. and Elias-Miro, J. and Reyimuaji, Y. and
                        Venturini, E.",
      title          = "{Novel measurements of anomalous triple gauge couplings
                        for the LHC}",
      journal        = "JHEP",
      volume         = "10",
      year           = "2017",
      pages          = "027",
      doi            = "10.1007/JHEP10(2017)027",
      eprint         = "1707.08060",
      archivePrefix  = "arXiv",
      primaryClass   = "hep-ph",
      reportNumber   = "SISSA-35-2017-FISI",
      SLACcitation   = "%%CITATION = ARXIV:1707.08060;%%"
}

@article{LUCID2,
  author={G. Avoni and others},
  title={The new LUCID-2 detector for luminosity measurement and monitoring in ATLAS},
  journal={JINST},
  volume={13},
  number={07},
  pages={P07017},
  doi="10.1088/1748-0221/13/07/P07017",
  year={2018},
}

@techreport{ATLAS-CONF-2019-021,
      author        = "{ATLAS Collaboration}", 
      title         = "{Luminosity determination in $pp$ collisions at
                       $\sqrt{s}=13$ TeV using the ATLAS detector at the LHC}",
      institution   = "CERN",
      collaboration = "ATLAS Collaboration",
      address       = "Geneva",
      number        = "ATLAS-CONF-2019-021",
      month         = "Jun",
      year          = "2019",
      reportNumber  = "ATLAS-CONF-2019-021",
      url           = "https://cds.cern.ch/record/2677054",
}

@misc{tensorflow2015-whitepaper,
	title={ {TensorFlow}: Large-Scale Machine Learning on Heterogeneous Systems},
	url={https://www.tensorflow.org/},
    eprint = "1603.04467",
    archivePrefix = "arXiv",
	doi = {10.48550/ARXIV.1603.04467},
	note={Software available from tensorflow.org},
	author={
	Martin~Abadi and
	Ashish~Agarwal and
	Paul~Barham and
	Eugene~Brevdo and
	Zhifeng~Chen and
	Craig~Citro and
	Greg~S.~Corrado and
	Andy~Davis and
	Jeffrey~Dean and
	Matthieu~Devin and
	Sanjay~Ghemawat and
	Ian~Goodfellow and
	Andrew~Harp and
	Geoffrey~Irving and
	Michael~Isard and
	Yangqing Jia and
	Rafal~Jozefowicz and
	Lukasz~Kaiser and
	Manjunath~Kudlur and
	Josh~Levenberg and
	Dandelion~Man\'{e} and
	Rajat~Monga and
	Sherry~Moore and
	Derek~Murray and
	Chris~Olah and
	Mike~Schuster and
	Jonathon~Shlens and
	Benoit~Steiner and
	Ilya~Sutskever and
	Kunal~Talwar and
	Paul~Tucker and
	Vincent~Vanhoucke and
	Vijay~Vasudevan and
	Fernanda~Vi\'{e}gas and
	Oriol~Vinyals and
	Pete~Warden and
	Martin~Wattenberg and
	Martin~Wicke and
	Yuan~Yu and
	Xiaoqiang~Zheng},
	year={2015},
}

@online{chollet2015keras,
	title={Keras},
	author={Chollet, Francois and others},
	year={2015},
	publisher={GitHub},
	url={https://github.com/fchollet/keras},
}

@article{Andreassen:2019nnm,
    author = "Andreassen, Anders and Nachman, Benjamin",
    title = "{Neural networks for full phase-space reweighting and parameter tuning}",
    eprint = "1907.08209",
    archivePrefix = "arXiv",
    primaryClass = "hep-ph",
    doi = "10.1103/PhysRevD.101.091901",
    journal = "Phys. Rev. D",
    volume = "101",
    number = "9",
    pages = "091901",
    year = "2020"
}

@techreport{ATL-PHYS-PUB-2016-005,
      author        = "{ATLAS Collaboration}",
      title         = "{Modelling of the $t\bar{t}H$ and $t\bar{t}V$ $(V=W,Z)$
                       processes for $\sqrt{s}=13$ TeV ATLAS analyses}",
      institution   = "CERN",
      collaboration = "ATLAS Collaboration",
      address       = "Geneva",
      number        = "ATL-PHYS-PUB-2016-005",
      month         = "Jan",
      year          = "2016",
      reportNumber  = "ATL-PHYS-PUB-2016-005",
      url           = "https://cds.cern.ch/record/2120826",
}

@article{WZVBS36fb,
    author = "{ATLAS Collaboration}",
    title = "{Observation of electroweak $W^{\pm}Z$ boson pair production in association with two jets in $pp$ collisions at $\sqrt{s} =$ 13 TeV with the ATLAS detector}",
    eprint = "1812.09740",
    archivePrefix = "arXiv",
    primaryClass = "hep-ex",
    reportNumber = "CERN-EP-2018-286",
    doi = "10.1016/j.physletb.2019.05.012",
    journal = "Phys. Lett. B",
    volume = "793",
    pages = "469",
    year = "2019"
}

@article{WZPol36fb,
    author = "{ATLAS Collaboration}",
    title = "{Measurement of $W^{\pm}Z$ production cross sections and gauge boson polarisation in $pp$ collisions at $\sqrt{s} = 13$ TeV with the ATLAS detector}",
    eprint = "1902.05759",
    archivePrefix = "arXiv",
    primaryClass = "hep-ex",
    reportNumber = "CERN-EP-2018-327",
    doi = "10.1140/epjc/s10052-019-7027-6",
    journal = "Eur. Phys. J. C",
    volume = "79",
    number = "6",
    pages = "535",
    year = "2019"
}

@article{BuarqueFranzosi:2019boy,
    author = "Buarque Franzosi, Diogo and Mattelaer, Olivier and Ruiz, Richard and Shil, Sujay",
    title = "{Automated predictions from polarized matrix elements}",
    eprint = "1912.01725",
    archivePrefix = "arXiv",
    primaryClass = "hep-ph",
    reportNumber = "CP3-19-52, VBSCan-PUB-09-19, MCnet-19-24, IP/BBSR/2019-10",
    doi = "10.1007/JHEP04(2020)082",
    journal = "JHEP",
    volume = "04",
    pages = "082",
    year = "2020"
}

@article{antikt,
      author         = "Cacciari, Matteo and Salam, Gavin P. and Soyez, Gregory",
      title          = "{The anti-$k_t$ jet clustering algorithm}",
      journal        = "JHEP",
      volume         = "04",
      year           = "2008",
      pages          = "063",
      doi            = "10.1088/1126-6708/2008/04/063",
      eprint         = "0802.1189",
      archivePrefix  = "arXiv",
      primaryClass   = "hep-ph",
      reportNumber   = "LPTHE-07-03",
      SLACcitation   = "%%CITATION = ARXIV:0802.1189;%%"
}

@Article{Fastjet,
      author         = "Cacciari, Matteo and Salam, Gavin P. and Soyez, Gregory",
      title          = "{FastJet user manual}",
      journal        = "Eur. Phys. J. C",
      volume         = "72",
      year           = "2012",
      pages          = "1896",
      doi            = "10.1140/epjc/s10052-012-1896-2",
      eprint         = "1111.6097",
      archivePrefix  = "arXiv",
      primaryClass   = "hep-ph",
      reportNumber   = "CERN-PH-TH-2011-297",
      SLACcitation   = "%%CITATION = ARXIV:1111.6097;%%"
}

@article{Malaescu:2011yg,
      author         = "Malaescu, Bogdan",
      title          = "{An Iterative, Dynamically Stabilized(IDS) Method of Data
                        Unfolding}",
      journal        = "{Proceedings of the PHYSTAT 2011 Workshop, CERN, Geneva, Switzerland}",
      xvolume         = "362",
      year           = "2011",
      pages          = "271",
      eprint         = "1106.3107",
      archivePrefix  = "arXiv",
      primaryClass   = "physics.data-an",
      reportNumber   = "CERN-2011-00"
 }

@article{Hirschi:2015iia,
    author = "Hirschi, Valentin and Mattelaer, Olivier",
    title = "{Automated event generation for loop-induced processes}",
    eprint = "1507.00020",
    archivePrefix = "arXiv",
    primaryClass = "hep-ph",
    reportNumber = "IPPP-15-35, DCPT-15-70, MCNET-15-14",
    doi = "10.1007/JHEP10(2015)146",
    journal = "JHEP",
    volume = "10",
    pages = "146",
    year = "2015"
}

@article{Aad:2020wog,
    author = "{ATLAS Collaboration}",
    collaboration = "ATLAS",
    title = "{Observation of the associated production of a top quark and a $Z$ boson in $pp$ collisions at $\sqrt{s} = 13$ TeV with the ATLAS detector}",
    eprint = "2002.07546",
    archivePrefix = "arXiv",
    primaryClass = "hep-ex",
    reportNumber = "CERN-EP-2019-273",
    doi = "10.1007/JHEP07(2020)124",
    journal = "JHEP",
    volume = "07",
    pages = "124",
    year = "2020"
}

@article{CMS:2021icx,
    author = "{CMS Collaboration}",
    collaboration = "CMS",
    title = "{Measurement of the inclusive and differential WZ production cross sections, polarization angles, and triple gauge couplings in pp collisions at $\sqrt{s}$ = 13 TeV}",
    eprint = "2110.11231",
    archivePrefix = "arXiv",
    primaryClass = "hep-ex",
    reportNumber = "CMS-SMP-20-014, CERN-EP-2021-163",
    month = "10",
    year = "2021"
}

@article{Baur:1994ia,
    author = "Baur, U. and Han, Tao and Ohnemus, J.",
    title = "{Amplitude zeros in W$^{+/-}$ Z production}",
    eprint = "hep-ph/9403248",
    archivePrefix = "arXiv",
    reportNumber = "FSU-HEP-940307, UCD-94-5",
    doi = "10.1103/PhysRevLett.72.3941",
    journal = "Phys. Rev. Lett.",
    volume = "72",
    pages = "3941",
    year = "1994"
}

@Article{PDG,
  author    = "Beringer, J. and others",
  title     = "{The Review of Particle Physics}",
  journal   = "Phys. Rev. D",
  volume    = "86",
  year      = "2012"
}

@article{CT10,
      author         = "Lai, Hung-Liang and Guzzi, Marco and Huston, Joey and Li,
                        Zhao and Nadolsky, Pavel M. and Pumplin, Jon and Yuan, C.
                        -P.",
      title          = "{New parton distributions for collider physics}",
      journal        = "Phys. Rev. D",
      volume         = "82",
      year           = "2010",
      pages          = "074024",
      doi            = "10.1103/PhysRevD.82.074024",
      eprint         = "1007.2241",
      archivePrefix  = "arXiv",
      primaryClass   = "hep-ph",
      reportNumber   = "MSUHEP-100707, SMU-HEP-10-10",
      SLACcitation   = "%%CITATION = ARXIV:1007.2241;%%"
}

@article{Aad:2010ah,
      author         = "{ATLAS Collaboration}",
      title          = "{The ATLAS Simulation Infrastructure}",
      collaboration  = "ATLAS",
      journal        = "Eur. Phys. J. C",
      volume         = "70",
      year           = "2010",
      pages          = "823",
      doi            = "10.1140/epjc/s10052-010-1429-9",
      eprint         = "1005.4568",
      archivePrefix  = "arXiv",
      primaryClass   = "physics.ins-det",
      SLACcitation   = "%%CITATION = ARXIV:1005.4568;%%"
}

@article{Aad:2014pda,
      author         = "{ATLAS Collaboration}",
      title          = "{Search for supersymmetry at $\sqrt{s}$=8 \TeV{} in final
                        states with jets and two same-sign leptons or three
                        leptons with the ATLAS detector}",
      collaboration  = "ATLAS",
      journal        = "JHEP",
      volume         = "06",
      year           = "2014",
      pages          = "035",
      doi            = "10.1007/JHEP06(2014)035",
      eprint         = "1404.2500",
      archivePrefix  = "arXiv",
      primaryClass   = "hep-ex",
      reportNumber   = "CERN-PH-EP-2014-044",
      SLACcitation   = "%%CITATION = ARXIV:1404.2500;%%"
}

@Article{Agostinelli:2002hh,
     author    = "Agostinelli, S. and others",
 collaboration = "GEANT4",
     title     = "{GEANT4 -- a simulation toolkit}",
     journal   = "Nucl. Instrum. Meth. A",
     volume    = "506",
     year      = "2003",
     pages     = "250",
     doi       = "10.1016/S0168-9002(03)01368-8",
     SLACcitation  = "%%CITATION = NUIMA,A506,250;%%"
}

@article{powheg,
      author         = "Frixione, Stefano and Nason, Paolo and Oleari, Carlo",
      title          = "{Matching NLO QCD computations with parton shower simulations: the POWHEG method}",
      journal        = "JHEP",
      volume         = "11",
      year           = "2007",
      pages          = "070",
      doi            = "10.1088/1126-6708/2007/11/070",
      eprint         = "0709.2092",
      archivePrefix  = "arXiv",
      primaryClass   = "hep-ph",
      reportNumber   = "BICOCCA-FT-07-9, GEF-TH-21-2007",
      SLACcitation   = "%%CITATION = ARXIV:0709.2092;%%"
}

@article{powheg1:2004,
      author         = "Nason, Paolo",
      title          = "{A New method for combining NLO QCD with shower Monte
                        Carlo algorithms}",
      journal        = "JHEP",
      volume         = "11",
      year           = "2004",
      pages          = "040",
      doi            = "10.1088/1126-6708/2004/11/040",
      eprint         = "hep-ph/0409146",
      archivePrefix  = "arXiv",
      reportNumber   = "BICOCCA-FT-04-11",
      SLACcitation   = "%%CITATION = HEP-PH/0409146;%%"
}

@article{Alioli:2010xd,
      author         = "Alioli, Simone and Nason, Paolo and Oleari, Carlo and Re,
                        Emanuele",
      title          = "{A general framework for implementing NLO calculations in
                        shower Monte Carlo programs: the POWHEG BOX}",
      journal        = "JHEP",
      volume         = "06",
      year           = "2010",
      pages          = "043",
      doi            = "10.1007/JHEP06(2010)043",
      eprint         = "1002.2581",
      archivePrefix  = "arXiv",
      primaryClass   = "hep-ph",
      reportNumber   = "DESY-10-018, SFB-CPP-10-22, IPPP-10-11, DCPT-10-22",
      SLACcitation   = "%%CITATION = ARXIV:1002.2581;%%"
}

@article{Ball:2014uwa,
      author         = "Ball, Richard D. and others",
      title          = "{Parton distributions for the LHC run II}",
      collaboration  = "NNPDF",
      journal        = "JHEP",
      volume         = "04",
      year           = "2015",
      pages          = "040",
      doi            = "10.1007/JHEP04(2015)040",
      eprint         = "1410.8849",
      archivePrefix  = "arXiv",
      primaryClass   = "hep-ph",
      reportNumber   = "EDINBURGH-2014-15, IFUM-1034-FT, CERN-PH-TH-2013-253,OUTP-14-11P, CAVENDISH-HEP-14-11",
      SLACcitation   = "%%CITATION = ARXIV:1410.8849;%%"
}

@article{PDF4LHC,
      author         = "Butterworth, Jon and others",
      title          = "{PDF4LHC recommendations for LHC Run II}",
      journal        = "J. Phys. G",
      volume         = "43",
      year           = "2016",
      pages          = "023001",
      doi            = "10.1088/0954-3899/43/2/023001",
      eprint         = "1510.03865",
      archivePrefix  = "arXiv",
      primaryClass   = "hep-ph",
      reportNumber   = "OUTP-15-17P, SMU-HEP-15-12, TIF-UNIMI-2015-14,
                        LCTS-2015-27, CERN-PH-TH-2015-249",
      SLACcitation   = "%%CITATION = ARXIV:1510.03865;%%"
}

@article{Pumplin:2002vw,
      author         = "Pumplin, J. and Stump, D. R. and Huston, J. and Lai, H.
                        L. and Nadolsky, Pavel M. and Tung, W. K.",
      title          = "{New Generation of Parton Distributions with
                        Uncertainties from Global QCD Analysis}",
      journal        = "JHEP",
      volume         = "07",
      year           = "2002",
      pages          = "012",
      doi            = "10.1088/1126-6708/2002/07/012",
      eprint         = "hep-ph/0201195",
      archivePrefix  = "arXiv",
      reportNumber   = "MSU-HEP-011101",
      SLACcitation   = "%%CITATION = HEP-PH/0201195;%%"
}

@article{Sjostrand:2007gs,
      author         = "Sj{\"o}strand, Torbjorn and Mrenna, Stephen and Skands, Peter",
      title          = "{A brief introduction to PYTHIA 8.1}",
      journal        = "Comput. Phys. Commun.",
      volume         = "178",
      year           = "2008",
      pages          = "852",
      doi            = "10.1016/j.cpc.2008.01.036",
      eprint         = "0710.3820",
      archivePrefix  = "arXiv",
      primaryClass   = "hep-ph",
      reportNumber   = "CERN-LCGAPP-2007-04, LU-TP-07-28,
                        FERMILAB-PUB-07-512-CD-T",
      SLACcitation   = "%%CITATION = ARXIV:0710.3820;%%"
}

@article{Melia:2011tj,
      author         = "Melia, Tom and Nason, Paolo and R{\"o}ntsch, Raoul and
                        Zanderighi, Giulia",
      title          = "{$W^{+}W^{-}$, $WZ$ and $ZZ$ production in the POWHEG BOX}",
      journal        = "JHEP",
      volume         = "11",
      year           = "2011",
      pages          = "078",
      doi            = "10.1007/JHEP11(2011)078",
      eprint         = "1107.5051",
      archivePrefix  = "arXiv",
      primaryClass   = "hep-ph",
      SLACcitation   = "%%CITATION = ARXIV:1107.5051;%%"
}

@article{Aad:2015ina,
      author         = "{ATLAS Collaboration}",
      title          = "{Performance of pile-up mitigation techniques for jets in
                        $pp$ collisions at $\sqrt{s}=8$  TeV using the ATLAS
                        detector}",
      collaboration  = "ATLAS",
      journal        = "Eur. Phys. J. C",
      volume         = "76",
      year           = "2016",
      number         = "11",
      pages          = "581",
      doi            = "10.1140/epjc/s10052-016-4395-z",
      eprint         = "1510.03823",
      archivePrefix  = "arXiv",
      primaryClass   = "hep-ex",
      reportNumber   = "CERN-PH-EP-2015-206",
      SLACcitation   = "%%CITATION = ARXIV:1510.03823;%%"
}

@article{ATLASWZ_8TeV,
      author         = "{ATLAS Collaboration}",
      title          = "{Measurements of $W^\pm Z$ production cross sections in
                        $pp$ collisions at $\sqrt{s} = 8$ TeV with the ATLAS
                        detector and limits on anomalous gauge boson
                        self-couplings}",
      collaboration  = "ATLAS",
      journal        = "Phys. Rev. D",
      volume         = "93",
      year           = "2016",
      pages          = "092004",
      doi            = "10.1103/PhysRevD.93.092004",
      eprint         = "1603.02151",
      archivePrefix  = "arXiv",
      primaryClass   = "hep-ex",
      reportNumber   = "CERN-EP-2016-017",
      SLACcitation   = "%%CITATION = ARXIV:1603.02151;%%"
}

@article{AZNLO:2014,
      author         = "{ATLAS Collaboration}",
      title          = "{Measurement of the $Z/\gamma^{\ast}$ boson transverse momentum distribution in pp collisions at $\sqrt{s} = 7$ \TeV{} with the ATLAS detector}",      
      journal        = "JHEP",
      volume         = "09",
      pages          = "145",
      doi            = "10.1007/JHEP09(2014)145",
      year           = "2014",
      eprint         = "1406.3660",
      archivePrefix  = "arXiv",
      primaryClass   = "hep-ex",
      reportNumber   = "CERN-PH-EP-2014-075",
      SLACcitation   = "%%CITATION = ARXIV:1406.3660;%%",      
}

@article{Ball:2012cx,
      author         = "Ball, Richard D. and others",
      collaboration  = "NNPDF Collaboration",
      title          = "{Parton distributions with LHC data}",
      journal        = "Nucl. Phys. B",
      volume         = "867",
      year           = "2013",
      pages          = "244",
      doi            = "10.1016/j.nuclphysb.2012.10.003",
      eprint         = "1207.1303",
      archivePrefix  = "arXiv",
      primaryClass   = "hep-ph",
      reportNumber   = "EDINBURGH-2012-08, IFUM-FT-997, FR-PHENO-2012-014,
                        RWTH-TTK-12-25, CERN-PH-TH-2012-037, SFB-CPP-12-47\,
                        --CERN-PH-TH-2012-037",
      SLACcitation   = "%%CITATION = ARXIV:1207.1303;%%"
}

@article{Grazzini:2017ckn,
      author         = "{M.~Grazzini, S.~Kallweit, D.~Rathlev and M.~Wiesemann}",
      title          = "{$W^\pm Z$ production at the LHC: fiducial cross sections
                        and distributions in NNLO QCD}",
      journal        = "JHEP",
      volume         = "05",
      year           = "2017",
      pages          = "139",
      doi            = "10.1007/JHEP05(2017)139",
      eprint         = "1703.09065",
      archivePrefix  = "arXiv",
      primaryClass   = "hep-ph",
      reportNumber   = "ZU-TH-06-17, CERN-TH-2017-065",
      SLACcitation   = "%%CITATION = ARXIV:1703.09065;%%"
}

@article{Cascioli:2011va,
      author         = "Cascioli, Fabio and Maierh{\"o}fer, Philipp and Pozzorini,
                        Stefano",
      title          = "{Scattering Amplitudes with Open Loops}",
      journal        = "Phys. Rev. Lett.",
      volume         = "108",
      year           = "2012",
      pages          = "111601",
      doi            = "10.1103/PhysRevLett.108.111601",
      eprint         = "1111.5206",
      archivePrefix  = "arXiv",
      primaryClass   = "hep-ph",
      reportNumber   = "ZU-TH-23-11, LPN11-66",
      SLACcitation   = "%%CITATION = ARXIV:1111.5206;%%"
}

@article{Mirkes:1994eb,
      author         = "Mirkes, E. and Ohnemus, J.",
      title          = "{$W$ and $Z$ polarization effects in hadronic
                        collisions}",
      journal        = "Phys. Rev. D",
      volume         = "50",
      year           = "1994",
      pages          = "5692",
      doi            = "10.1103/PhysRevD.50.5692",
      eprint         = "hep-ph/9406381",
      archivePrefix  = "arXiv",
      reportNumber   = "MAD-PH-834, UCD-94-23",
      SLACcitation   = "%%CITATION = HEP-PH/9406381;%%"
}

@article{Groote:2012xr,
      author         = "Groote, S. and Korner, J. G. and Tuvike, P.",
      title          = "{$O(\alpha_s)$ corrections to the decays of polarized
                        $W^\pm$ and $Z$ bosons into massive quark pairs}",
      journal        = "Eur. Phys. J. C",
      volume         = "72",
      year           = "2012",
      pages          = "2177",
      doi            = "10.1140/epjc/s10052-012-2177-9",
      eprint         = "1204.5295",
      archivePrefix  = "arXiv",
      primaryClass   = "hep-ph",
      reportNumber   = "MZ-TH-12-16",
      SLACcitation   = "%%CITATION = ARXIV:1204.5295;%%"
}

@article{Stirling:2012zt,
      author         = "Stirling, W. J. and Vryonidou, E.",
      title          = "{Electroweak gauge boson polarisation at the LHC}",
      journal        = "JHEP",
      volume         = "07",
      year           = "2012",
      pages          = "124",
      doi            = "10.1007/JHEP07(2012)124",
      eprint         = "1204.6427",
      archivePrefix  = "arXiv",
      primaryClass   = "hep-ph",
      SLACcitation   = "%%CITATION = ARXIV:1204.6427;%%"
}

@article{Gounaris:1992kp,
      author         = "Gounaris, G. and Layssac, J. and Moultaka, G. and Renard,
                        F. M.",
      title          = "{Analytic expressions of cross-sections, asymmetries and
                        W density matrices for e+ e- ---> W+ W- with general three
                        boson couplings}",
      journal        = "Int. J. Mod. Phys. A",
      volume         = "8",
      year           = "1993",
      pages          = "3285",
      doi            = "10.1142/S0217751X93001326",
      reportNumber   = "PM-92-37, THES-TP-92-15",
      SLACcitation   = "%%CITATION = IMPAE,A8,3285;%%"
}

@article{Aaboud:2018tkc,
      author         = "{ATLAS Collaboration}",
      title          = "{Performance of missing transverse momentum
                        reconstruction with the ATLAS detector using proton-proton
                        collisions at $\sqrt{s}$ = 13 TeV}",
      collaboration  = "ATLAS",
      journal        = "Eur. Phys. J. C",
      volume         = "78",
      year           = "2018",
      number         = "11",
      pages          = "903",
      doi            = "10.1140/epjc/s10052-018-6288-9",
      eprint         = "1802.08168",
      archivePrefix  = "arXiv",
      primaryClass   = "hep-ex",
      reportNumber   = "CERN-EP-2017-274",
      SLACcitation   = "%%CITATION = ARXIV:1802.08168;%%"
}

@article{Aaltonen:2012rz,
      author         = "{CDF and D0 Collaborations}",
      title          = "{Combination of CDF and D0 measurements of the $W$ boson
                        helicity in top quark decays}",
      collaboration  = "CDF, D0",
      journal        = "Phys. Rev. D",
      volume         = "85",
      year           = "2012",
      pages          = "071106",
      doi            = "10.1103/PhysRevD.85.071106",
      eprint         = "1202.5272",
      archivePrefix  = "arXiv",
      primaryClass   = "hep-ex",
      reportNumber   = "SLAC-PUB-14955, FERMILAB-PUB-12-043-E",
      SLACcitation   = "%%CITATION = ARXIV:1202.5272;%%"
}

@article{Aaltonen:2010ha,
      author         = "{CDF Collaboration}",
      title          = "{Measurement of W-Boson Polarization in Top-Quark Decay
                        in $p\bar{p}$ Collisions at $\sqrt{s} = 1.96$ TeV}",
      collaboration  = "CDF",
      journal        = "Phys. Rev. Lett.",
      volume         = "105",
      year           = "2010",
      pages          = "042002",
      doi            = "10.1103/PhysRevLett.105.042002",
      eprint         = "1003.0224",
      archivePrefix  = "arXiv",
      primaryClass   = "hep-ex",
      reportNumber   = "FERMILAB-PUB-10-041-E",
      SLACcitation   = "%%CITATION = ARXIV:1003.0224;%%"
}

@article{Abazov:2010jn,
      author         = "{D0 Collaboration}",
      title          = "{Measurement of the W boson helicity in top quark decays
                        using 5.4 fb$^{-1}$ of $p\bar{p}$ collision data}",
      xcollaboration  = "D0",
      journal        = "Phys. Rev. D",
      volume         = "83",
      year           = "2011",
      pages          = "032009",
      doi            = "10.1103/PhysRevD.83.032009",
      eprint         = "1011.6549",
      archivePrefix  = "arXiv",
      primaryClass   = "hep-ex",
      reportNumber   = "FERMILAB-PUB-10-484-E",
      SLACcitation   = "%%CITATION = ARXIV:1011.6549;%%"
}

@article{Aad:2012ky,
      author         = "{ATLAS Collaboration}",
      title          = "{Measurement of the W boson polarization in top quark
                        decays with the ATLAS detector}",
      journal        = "JHEP",
      volume         = "06",
      year           = "2012",
      pages          = "088",
      doi            = "10.1007/JHEP06(2012)088",
      eprint         = "1205.2484",
      archivePrefix  = "arXiv",
      primaryClass   = "hep-ex",
      reportNumber   = "CERN-PH-EP-2012-112",
      SLACcitation   = "%%CITATION = ARXIV:1205.2484;%%"
}

@article{Khachatryan:2016fky,
      author         = "{CMS Collaboration}",
      title          = "{Measurement of the W boson helicity fractions in the
                        decays of top quark pairs to lepton $+$ jets final states
                        produced in pp collisions at $\sqrt s=$ 8TeV}",
      journal        = "Phys. Lett. B",
      volume         = "762",
      year           = "2016",
      pages          = "512",
      doi            = "10.1016/j.physletb.2016.10.007",
      eprint         = "1605.09047",
      archivePrefix  = "arXiv",
      primaryClass   = "hep-ex",
      reportNumber   = "CMS-TOP-13-008, CERN-EP-2016-118",
      SLACcitation   = "%%CITATION = ARXIV:1605.09047;%%"
}

@article{Chatrchyan:2011ig,
      author         = "{CMS Collaboration}",
      title          = "{Measurement of the Polarization of W Bosons with Large
                        Transverse Momenta in W+jets Events at the LHC}",
      collaboration  = "CMS",
      journal        = "Phys. Rev. Lett.",
      volume         = "107",
      year           = "2011",
      pages          = "021802",
      doi            = "10.1103/PhysRevLett.107.021802",
      eprint         = "1104.3829",
      archivePrefix  = "arXiv",
      primaryClass   = "hep-ex",
      reportNumber   = "CERN-PH-EP-2011-043, CMS-EWK-10-014",
      SLACcitation   = "%%CITATION = ARXIV:1104.3829;%%"
}

@article{Aaltonen:2011nr,
      author         = "{CDF Collaboration}",
      title          = "{First Measurement of the Angular Coefficients of
                        Drell-Yan $e^{+}e^{-}$ Pairs in the Z Mass Region from
                        $p\bar{p}$ Collisions at $\sqrt{s}$ = 1.96 TeV}",
      collaboration  = "CDF",
      journal        = "Phys. Rev. Lett.",
      volume         = "106",
      year           = "2011",
      pages          = "241801",
      doi            = "10.1103/PhysRevLett.106.241801",
      eprint         = "1103.5699",
      archivePrefix  = "arXiv",
      primaryClass   = "hep-ex",
      reportNumber   = "FERMILAB-PUB-11-153-E",
      SLACcitation   = "%%CITATION = ARXIV:1103.5699;%%"
}

@article{Khachatryan:2015paa,
      author         = "{CMS Collaboration}",
      title          = "{Angular coefficients of Z bosons produced in pp
                        collisions at $\sqrt{s}$ = 8 TeV and decaying to $\mu^+
                        \mu^-$ as a function of transverse momentum and rapidity}",
      collaboration  = "CMS",
      journal        = "Phys. Lett. B",
      volume         = "750",
      year           = "2015",
      pages          = "154",
      doi            = "10.1016/j.physletb.2015.08.061",
      eprint         = "1504.03512",
      archivePrefix  = "arXiv",
      primaryClass   = "hep-ex",
      reportNumber   = "CMS-SMP-13-010, CERN-PH-EP-2015-046",
      SLACcitation   = "%%CITATION = ARXIV:1504.03512;%%"
}

@article{Aad:2016izn,
      author         = "{ATLAS Collaboration}",
      title          = "{Measurement of the angular coefficients in $Z$-boson
                        events using electron and muon pairs from data taken at
                        $\sqrt{s}=8$ TeV with the ATLAS detector}",
      journal        = "JHEP",
      volume         = "08",
      year           = "2016",
      pages          = "159",
      doi            = "10.1007/JHEP08(2016)159",
      eprint         = "1606.00689",
      archivePrefix  = "arXiv",
      primaryClass   = "hep-ex",
      reportNumber   = "CERN-EP-2016-087",
      SLACcitation   = "%%CITATION = ARXIV:1606.00689;%%"
}

@article{Achard:2002bv,
      author         = "{L3 Collaboration}",
      title          = "{Measurement of W polarisation at LEP}",
      collaboration  = "L3",
      journal        = "Phys. Lett. B",
      volume         = "557",
      year           = "2003",
      pages          = "147",
      doi            = "10.1016/S0370-2693(03)00172-2",
      eprint         = "hep-ex/0301027",
      archivePrefix  = "arXiv",
      reportNumber   = "CERN-EP-2002-092",
      SLACcitation   = "%%CITATION = HEP-EX/0301027;%%"
}

@article{Aaboud:2016leb,
      author         = "{ATLAS Collaboration}",
      title          = "{Performance of the ATLAS trigger system in 2015}",
      collaboration  = "ATLAS",
      journal        = "Eur. Phys. J. C",
      volume         = "77",
      year           = "2017",
      number         = "5",
      pages          = "317",
      doi            = "10.1140/epjc/s10052-017-4852-3",
      eprint         = "1611.09661",
      archivePrefix  = "arXiv",
      primaryClass   = "hep-ex",
      reportNumber   = "CERN-EP-2016-241",
      SLACcitation   = "%%CITATION = ARXIV:1611.09661;%%"
}

@article{Aaboud:2017jcu,
      author         = "{ATLAS Collaboration}",
      title          = "{Jet energy scale measurements and their systematic
                        uncertainties in proton-proton collisions at $\sqrt{s} =
                        13$ TeV with the ATLAS detector}",
      collaboration  = "ATLAS",
      journal        = "Phys. Rev. D",
      volume         = "96",
      year           = "2017",
      number         = "7",
      pages          = "072002",
      doi            = "10.1103/PhysRevD.96.072002",
      eprint         = "1703.09665",
      archivePrefix  = "arXiv",
      primaryClass   = "hep-ex",
      reportNumber   = "CERN-EP-2017-038",
      SLACcitation   = "%%CITATION = ARXIV:1703.09665;%%"
}

@article{Abbiendi:2003wv,
      author         = "{OPAL Collaboration}",
      title          = "{W boson polarization at LEP2}",
      collaboration  = "OPAL",
      journal        = "Phys. Lett. B",
      volume         = "585",
      year           = "2004",
      pages          = "223-236",
      doi            = "10.1016/j.physletb.2004.02.003",
      eprint         = "hep-ex/0312047",
      archivePrefix  = "arXiv",
      reportNumber   = "CERN-EP-2003-088",
      SLACcitation   = "%%CITATION = HEP-EX/0312047;%%"
}

@article{Abdallah:2008sf,
      author         = "{DELPHI Collaboration}",
      title          = "{Study of W-boson polarisations and triple gauge boson
                        couplings in the reaction e+e- ---> W+W- at LEP 2}",
      collaboration  = "DELPHI",
      journal        = "Eur. Phys. J. C",
      volume         = "54",
      year           = "2008",
      pages          = "345-364",
      doi            = "10.1140/epjc/s10052-008-0528-3",
      eprint         = "0801.1235",
      archivePrefix  = "arXiv",
      primaryClass   = "hep-ex",
      reportNumber   = "CERN-PH-EP-2007-036",
      SLACcitation   = "%%CITATION = ARXIV:0801.1235;%%"
}

@article{Aaron:2009wp,
      author         = "{H1 Collaboration}",
      title          = "{Events with isolated leptons and missing transverse
                        momentum and measurement of $W$ production at HERA}",
      collaboration  = "H1",
      journal        = "Eur. Phys. J. C",
      volume         = "64",
      year           = "2009",
      pages          = "251-271",
      doi            = "10.1140/epjc/s10052-009-1160-6",
      eprint         = "0901.0488",
      archivePrefix  = "arXiv",
      primaryClass   = "hep-ex",
      reportNumber   = "DESY-08-170",
      SLACcitation   = "%%CITATION = ARXIV:0901.0488;%%"
}

@article{Panico:2017frx,
      author         = "Panico, Giuliano and Riva, Francesco and Wulzer, Andrea",
      title          = "{Diboson interference resurrection}",
      journal        = "Phys. Lett. B",
      volume         = "776",
      year           = "2018",
      pages          = "473-480",
      doi            = "10.1016/j.physletb.2017.11.068",
      eprint         = "1708.07823",
      archivePrefix  = "arXiv",
      primaryClass   = "hep-ph",
      reportNumber   = "CERN-TH-2017-185",
      SLACcitation   = "%%CITATION = ARXIV:1708.07823;%%"
}

@article{Sjostrand:2014zea,
      author         = "Sj{\"o}strand, Torbjorn and Ask, Stefan and Christiansen,
                        Jesper R. and Corke, Richard and Desai, Nishita and Ilten,
                        Philip and Mrenna, Stephen and Prestel, Stefan and
                        Rasmussen, Christine O. and Skands, Peter Z.",
      title          = "{An introduction to \textsc{PYTHIA} 8.2}",
      journal        = "Comput. Phys. Commun.",
      volume         = "191",
      year           = "2015",
      pages          = "159-177",
      doi            = "10.1016/j.cpc.2015.01.024",
      eprint         = "1410.3012",
      archivePrefix  = "arXiv",
      primaryClass   = "hep-ph",
      reportNumber   = "LU-TP-14-36, MCNET-14-22, CERN-PH-TH-2014-190,
                        FERMILAB-PUB-14-316-CD, DESY-14-178, SLAC-PUB-16122",
      SLACcitation   = "%%CITATION = ARXIV:1410.3012;%%"
}

@article{Aaboud:2017rwm,
      author         = "{ATLAS Collaboration}",
      title          = "{$ZZ \to \ell^{+}\ell^{-}\ell^{\prime +}\ell^{\prime -}$
                        cross-section measurements and search for anomalous triple
                        gauge couplings in 13 TeV $pp$ collisions with the ATLAS
                        detector}",
      collaboration  = "ATLAS",
      journal        = "Phys. Rev. D",
      volume         = "97",
      year           = "2018",
      number         = "3",
      pages          = "032005",
      doi            = "10.1103/PhysRevD.97.032005",
      eprint         = "1709.07703",
      archivePrefix  = "arXiv",
      primaryClass   = "hep-ex",
      reportNumber   = "CERN-EP-2017-163",
      SLACcitation   = "%%CITATION = ARXIV:1709.07703;%%"
}

@article{PhysRevD.98.030001,
  author = "{Particle Data Group}",
  title = "{Review of Particle Physics}",
  collaboration = "Particle Data Group",
  journal = "Phys. Rev. D",
  volume = "98",
  %issue = "3",
  pages = "030001",
  numpages = "1898",
  year = "2018",
  month = "Aug",
  publisher = "American Physical Society",
  doi = "10.1103/PhysRevD.98.030001",
  %url = "https://link.aps.org/doi/10.1103/PhysRevD.98.030001"
}

@article{Aaboud:2016mmw,
      author         = "{ATLAS Collaboration}",
      title          = "{Measurement of the Inelastic Proton-Proton Cross Section
                        at $\sqrt{s} = 13$ TeV with the ATLAS Detector at the
                        LHC}",
      collaboration  = "ATLAS",
      journal        = "Phys. Rev. Lett.",
      volume         = "117",
      year           = "2016",
      number         = "18",
      pages          = "182002",
      doi            = "10.1103/PhysRevLett.117.182002",
      eprint         = "1606.02625",
      archivePrefix  = "arXiv",
      primaryClass   = "hep-ex",
      reportNumber   = "CERN-EP-2016-140",
      SLACcitation   = "%%CITATION = ARXIV:1606.02625;%%"
}

@article{Franceschini:2017xkh,
    author = "Franceschini, Roberto and Panico, Giuliano and Pomarol, Alex and Riva, Francesco and Wulzer, Andrea",
    title = "{Electroweak precision tests in high-energy diboson processes}",
    eprint = "1712.01310",
    archivePrefix = "arXiv",
    primaryClass = "hep-ph",
    reportNumber = "CERN-TH-2017-252, RM3-TH-17-1",
    doi = "10.1007/JHEP02(2018)111",
    journal = "JHEP",
    volume = "02",
    pages = "111",
    year = "2018"
}

@article{Lonnblad:2001iq,
    author = "L{\"o}nnblad, Leif",
    title = "{Correcting the Colour-Dipole Cascade Model with Fixed Order Matrix Elements}",
    eprint = "hep-ph/0112284",
    archivePrefix = "arXiv",
    reportNumber = "LU-TP-01-38",
    doi = "10.1088/1126-6708/2002/05/046",
    journal = "JHEP",
    volume = "05",
    pages = "046",
    year = "2002"
}

@article{Lonnblad:2011xx,
    author = "L{\"o}nnblad, Leif and Prestel, Stefan",
    title = "{Matching tree-level matrix elements with interleaved showers}",
    eprint = "1109.4829",
    archivePrefix = "arXiv",
    primaryClass = "hep-ph",
    doi = "10.1007/JHEP03(2012)019",
    journal = "JHEP",
    volume = "03",
    pages = "019",
    year = "2012"
}

@article{ATLAS:2019qmc,
    author = "{ATLAS Collaboration}",
    collaboration = "ATLAS",
    title = "{Electron and photon performance measurements with the ATLAS detector using the 2015\textendash{}2017 LHC proton-proton collision data}",
    eprint = "1908.00005",
    archivePrefix = "arXiv",
    primaryClass = "hep-ex",
    reportNumber = "CERN-EP-2019-145",
    doi = "10.1088/1748-0221/14/12/P12006",
    journal = "JINST",
    volume = "14",
    number = "12",
    pages = "P12006",
    year = "2019"
}

@article{ATLAS:2020auj,
    author = "{ATLAS Collaboration}",
    collaboration = "ATLAS",
    title = "{Muon reconstruction and identification efficiency in ATLAS using the full Run 2 $pp$ collision data set at $\sqrt{s}=13$ TeV}",
    eprint = "2012.00578",
    archivePrefix = "arXiv",
    primaryClass = "hep-ex",
    reportNumber = "CERN-EP-2020-199",
    doi = "10.1140/epjc/s10052-021-09233-2",
    journal = "Eur. Phys. J. C",
    volume = "81",
    number = "7",
    pages = "578",
    year = "2021"
}

@article{ATLAS:2016lqx,
    author = "{ATLAS Collaboration}",
    collaboration = "ATLAS",
    title = "{Muon reconstruction performance of the ATLAS detector in proton\textendash{}proton collision data at $\sqrt{s}$ =13 TeV}",
    eprint = "1603.05598",
    archivePrefix = "arXiv",
    primaryClass = "hep-ex",
    reportNumber = "CERN-EP-2016-033",
    doi = "10.1140/epjc/s10052-016-4120-y",
    journal = "Eur. Phys. J. C",
    volume = "76",
    number = "5",
    pages = "292",
    year = "2016"
}

@article{ATLAS:2017ghe,
    author = "{ATLAS Collaboration}",
    collaboration = "ATLAS",
    title = "{Jet reconstruction and performance using particle flow with the ATLAS Detector}",
    eprint = "1703.10485",
    archivePrefix = "arXiv",
    primaryClass = "hep-ex",
    reportNumber = "CERN-EP-2017-024",
    doi = "10.1140/epjc/s10052-017-5031-2",
    journal = "Eur. Phys. J. C",
    volume = "77",
    number = "7",
    pages = "466",
    year = "2017"
}

@article{ATLAS:2020cli,
    author = "{ATLAS Collaboration}",
    collaboration = "ATLAS",
    title = "{Jet energy scale and resolution measured in proton\textendash{}proton collisions at $\sqrt{s}=13$~TeV with the ATLAS detector}",
    eprint = "2007.02645",
    archivePrefix = "arXiv",
    primaryClass = "hep-ex",
    reportNumber = "CERN-EP-2020-083",
    doi = "10.1140/epjc/s10052-021-09402-3",
    journal = "Eur. Phys. J. C",
    volume = "81",
    number = "8",
    pages = "689",
    year = "2021"
}

@techreport{ATL-PHYS-PUB-2017-003,
      author        = "{ATLAS Collaboration}",
      title         = "{Identification of Jets Containing $b$-Hadrons with
                       Recurrent Neural Networks at the ATLAS Experiment}",
      institution   = "CERN",
      collaboration = "ATLAS Collaboration",
      address       = "Geneva",
      reportNumber  = "ATL-PHYS-PUB-2017-003",
	  number        = "ATL-PHYS-PUB-2017-003",
      month         = "Mar",
      year          = "2017",
      url           = "https://cds.cern.ch/record/2255226",
}

@article{ATLAS:2019bwq,
    author = "{ATLAS Collaboration}",
    collaboration = "ATLAS",
    title = "{ATLAS b-jet identification performance and efficiency measurement with $t{\bar{t}}$ events in pp collisions at $\sqrt{s}=13$ TeV}",
    eprint = "1907.05120",
    archivePrefix = "arXiv",
    primaryClass = "hep-ex",
    reportNumber = "CERN-EP-2019-132",
    doi = "10.1140/epjc/s10052-019-7450-8",
    journal = "Eur. Phys. J. C",
    volume = "79",
    number = "11",
    pages = "970",
    year = "2019"
}

@techreport{ATL-PHYS-PUB-2017-013,
      author        = "{ATLAS Collaboration}",
      title         = "{Optimisation and performance studies of the ATLAS
                       $b$-tagging algorithms for the 2017-18 LHC run}",
      institution   = "CERN",
      collaboration = "ATLAS Collaboration",
      address       = "Geneva",
      reportNumber  = "ATL-PHYS-PUB-2017-013",
      number  = "ATL-PHYS-PUB-2017-013",
      month         = "Jul",
      year          = "2017",
      url           = "https://cds.cern.ch/record/2273281"
}

@article{Le:2022lrp,
    author = "Le, Duc Ninh and Baglio, Julien",
    title = "{Doubly-polarized $WZ$ hadronic cross sections at NLO QCD+EW accuracy}",
    eprint = "2203.01470",
    archivePrefix = "arXiv",
    primaryClass = "hep-ph",
    reportNumber = "CERN-TH-2022-027",
    month = "3",
    year = "2022"
}

@article{OPAL:2000wbs,
    author = "{OPAL Collaboration}",
    collaboration = "OPAL",
    title = "{Measurement of $W$ boson polarizations and CP violating triple gauge couplings from $W^{+} W^{-}$ production at LEP}",
    eprint = "hep-ex/0009021",
    archivePrefix = "arXiv",
    reportNumber = "CERN-EP-2000-113",
    doi = "10.1007/s100520100602",
    journal = "Eur. Phys. J. C",
    volume = "19",
    pages = "229--240",
    year = "2001"
}

@article{L3:2003tnr,
    author = "{L3 Collaboration}",
    collaboration = "L3",
    title = "{Study of spin and decay-plane correlations of $W$ bosons in the $e^{+} e^{-} \to W^{+} W^{-}$ process at LEP}",
    eprint = "hep-ex/0501036",
    archivePrefix = "arXiv",
    reportNumber = "CERN-EP-2003-072",
    doi = "10.1140/epjc/s2005-02137-9",
    journal = "Eur. Phys. J. C",
    volume = "40",
    pages = "333",
    year = "2005"
}

@article{DELPHI:2009wdg,
    author = "{DELPHI Collaboration}",
    collaboration = "DELPHI",
    title = "{Correlations between Polarisation States of W Particles in the Reaction e- e+ ---\ensuremath{>} W- W+ at LEP2 Energies 189-GeV - 209-GeV}",
    eprint = "0908.1023",
    archivePrefix = "arXiv",
    primaryClass = "hep-ex",
    reportNumber = "CERN-PH-EP-2009-003",
    doi = "10.1140/epjc/s10052-009-1123-y",
    journal = "Eur. Phys. J. C",
    volume = "63",
    pages = "611",
    year = "2009"
}

@article{ATLAS:2021fzm,
    author = "{ATLAS Collaboration}",
    collaboration = "ATLAS",
    title = "{Measurements of the inclusive and differential production cross sections of a top-quark\textendash{}antiquark pair in association with a Z~boson at $\sqrt{s} = 13$~TeV with the ATLAS detector}",
    eprint = "2103.12603",
    archivePrefix = "arXiv",
    primaryClass = "hep-ex",
    reportNumber = "CERN-EP-2021-01",
    doi = "10.1140/epjc/s10052-021-09439-4",
    journal = "Eur. Phys. J. C",
    volume = "81",
    number = "8",
    pages = "737",
    year = "2021"
}

@article{ATLAS:2019dpa,
    author = "{ATLAS Collaboration}",
    collaboration = "ATLAS",
    title = "{Performance of electron and photon triggers in ATLAS during LHC Run 2}",
    eprint = "1909.00761",
    archivePrefix = "arXiv",
    primaryClass = "hep-ex",
    reportNumber = "CERN-EP-2019-169",
    doi = "10.1140/epjc/s10052-019-7500-2",
    journal = "Eur. Phys. J. C",
    volume = "80",
    number = "1",
    pages = "47",
    year = "2020"
}

@article{ATLAS:2020gty,
    author = "{ATLAS Collaboration}",
    collaboration = "ATLAS",
    title = "{Performance of the ATLAS muon triggers in Run 2}",
    eprint = "2004.13447",
    archivePrefix = "arXiv",
    primaryClass = "physics.ins-det",
    reportNumber = "CERN-EP-2020-031",
    doi = "10.1088/1748-0221/15/09/p09015",
    journal = "JINST",
    volume = "15",
    number = "09",
    pages = "P09015",
    year = "2020"
}

@techreport{ATL-PHYS-PUB-2016-017,
    author = "{ATLAS Collaboration}",
      title         = "{The Pythia 8 A3 tune description of ATLAS minimum bias
                       and inelastic measurements incorporating the
                       Donnachie-Landshoff diffractive model}",
      institution   = "CERN",
      collaboration = "ATLAS Collaboration",
      address       = "Geneva",
      reportNumber  = "ATL-PHYS-PUB-2016-017",
      month         = "Aug",
      year          = "2016",
      url           = "https://cds.cern.ch/record/2206965",
}

@article{Frederix:2012ps,
    author = "Frederix, Rikkert and Frixione, Stefano",
    title = "{Merging meets matching in MC@NLO}",
    eprint = "1209.6215",
    archivePrefix = "arXiv",
    primaryClass = "hep-ph",
    reportNumber = "CERN-PH-TH-2012-247, ZU-TH-21-12",
    doi = "10.1007/JHEP12(2012)061",
    journal = "JHEP",
    volume = "12",
    pages = "061",
    year = "2012"
}

@article{Sherpa:2019gpd,
    author = "Bothmann, Enrico and others",
    collaboration = "Sherpa",
    title = "{Event Generation with Sherpa 2.2}",
    eprint = "1905.09127",
    archivePrefix = "arXiv",
    primaryClass = "hep-ph",
    reportNumber = "FERMILAB-PUB-19-218-T, SLAC-PUB-17433, IPPP/19/42, MCNET-19-11",
    doi = "10.21468/SciPostPhys.7.3.034",
    journal = "SciPost Phys.",
    volume = "7",
    number = "3",
    pages = "034",
    year = "2019"
}

@article{Bierlich:2019rhm,
    author = "Bierlich, Christian and others",
    title = "{Robust Independent Validation of Experiment and Theory: Rivet version 3}",
    eprint = "1912.05451",
    archivePrefix = "arXiv",
    primaryClass = "hep-ph",
    reportNumber = "MCnet-19-26",
    doi = "10.21468/SciPostPhys.8.2.026",
    journal = "SciPost Phys.",
    volume = "8",
    pages = "026",
    year = "2020"
}

@article{Maguire:2017ypu,
    author = "Maguire, Eamonn and Heinrich, Lukas and Watt, Graeme",
    editor = "Mount, Richard and Tull, Craig",
    title = "{HEPData: a repository for high energy physics data}",
    eprint = "1704.05473",
    archivePrefix = "arXiv",
    primaryClass = "hep-ex",
    reportNumber = "IPPP-17-31",
    doi = "10.1088/1742-6596/898/10/102006",
    journal = "J. Phys. Conf. Ser.",
    volume = "898",
    number = "10",
    pages = "102006",
    year = "2017"
}

@techreport{Cranmer:1456844,
      author        = "Cranmer, Kyle and Lewis, George and Moneta, Lorenzo and
                       Shibata, Akira and Verkerke, Wouter",
      title         = "{HistFactory: A tool for creating statistical models for
                       use with RooFit and RooStats}",
      institution   = "New York U.",
      collaboration = "ROOT Collaboration",
      address       = "New York",
      reportNumber  = "CERN-OPEN-2012-016",
      year          = "2012",
      url           = "https://cds.cern.ch/record/1456844",
}

@article{ATLAS:2022swp,
	author = "{ATLAS Collaboration}",
    collaboration = "ATLAS",
    title = "{Tools for estimating fake/non-prompt lepton backgrounds with the ATLAS detector at the LHC}",
    eprint = "2211.16178",
    archivePrefix = "arXiv",
    primaryClass = "hep-ex",
    reportNumber = "CERN-EP-2022-214",
    journal   = "submitted to JINST.",
    month = "11",
    year = "2022"
}

@Article{PERF-2007-01,
    author         = "{ATLAS Collaboration}",
    title          = "{The ATLAS Experiment at the CERN Large Hadron Collider}",
    journal        = "JINST",
    volume         = "3",
    year           = "2008",
    pages          = "S08003",
    doi            = "10.1088/1748-0221/3/08/S08003",
    primaryClass   = "hep-ex",
}

@Article{STDM-2022-01,
    author         = "{ATLAS Collaboration}",
    title          = "{Observation of gauge boson joint-polarisation states in \(W^{\pm}Z\) production from \(pp\) collisions at \(\sqrt{s} = 13\,\text{TeV}\) with the ATLAS detector}",
    year           = "2022",
    reportNumber   = "CERN-EP-2022-202",
    eprint         = "2211.09435",
    archivePrefix  = "arXiv",
    primaryClass   = "hep-ex",
}

@Booklet{ATL-SOFT-PUB-2021-001,
    author         = "{ATLAS Collaboration}",
    title          = "{The ATLAS Collaboration Software and Firmware}",
    howpublished   = "{ATL-SOFT-PUB-2021-001}",
    url            = "https://cds.cern.ch/record/2767187",
    year           = "2021",
}

@Booklet{ATL-SOFT-PUB-2021-003,
    author         = "{ATLAS Collaboration}",
    title          = "{ATLAS Computing Acknowledgements}",
    howpublished   = "{ATL-SOFT-PUB-2021-003}",
    url            = "https://cds.cern.ch/record/2776662",
    year           = "2021",
}
 
\clearpage
 
\begin{flushleft}
\hypersetup{urlcolor=black}
{\Large The ATLAS Collaboration}

\bigskip

\AtlasOrcid[0000-0002-6665-4934]{G.~Aad}$^\textrm{\scriptsize 101}$,
\AtlasOrcid[0000-0002-5888-2734]{B.~Abbott}$^\textrm{\scriptsize 119}$,
\AtlasOrcid[0000-0002-7248-3203]{D.C.~Abbott}$^\textrm{\scriptsize 102}$,
\AtlasOrcid[0000-0002-1002-1652]{K.~Abeling}$^\textrm{\scriptsize 55}$,
\AtlasOrcid[0000-0002-8496-9294]{S.H.~Abidi}$^\textrm{\scriptsize 29}$,
\AtlasOrcid[0000-0002-9987-2292]{A.~Aboulhorma}$^\textrm{\scriptsize 35e}$,
\AtlasOrcid[0000-0001-5329-6640]{H.~Abramowicz}$^\textrm{\scriptsize 150}$,
\AtlasOrcid[0000-0002-1599-2896]{H.~Abreu}$^\textrm{\scriptsize 149}$,
\AtlasOrcid[0000-0003-0403-3697]{Y.~Abulaiti}$^\textrm{\scriptsize 116}$,
\AtlasOrcid[0000-0003-0762-7204]{A.C.~Abusleme~Hoffman}$^\textrm{\scriptsize 136a}$
% ...
\bigskip
\\

% ...
$^{29}$Physics Department, Brookhaven National Laboratory, Upton NY; United States of America.\\
% ...
$^{35}$$^{(a)}$Facult\'e des Sciences Ain Chock, R\'eseau Universitaire de Physique des Hautes Energies - Universit\'e Hassan II, Casablanca;$^{(b)}$Facult\'{e} des Sciences, Universit\'{e} Ibn-Tofail, K\'{e}nitra;$^{(c)}$Facult\'e des Sciences Semlalia, Universit\'e Cadi Ayyad, LPHEA-Marrakech;$^{(d)}$LPMR, Facult\'e des Sciences, Universit\'e Mohamed Premier, Oujda;$^{(e)}$Facult\'e des sciences, Universit\'e Mohammed V, Rabat;$^{(f)}$Institute of Applied Physics, Mohammed VI Polytechnic University, Ben Guerir; Morocco.\\
% ...
$^{55}$II. Physikalisches Institut, Georg-August-Universit\"{a}t G\"ottingen, G\"ottingen; Germany.\\
% ...
$^{101}$CPPM, Aix-Marseille Universit\'e, CNRS/IN2P3, Marseille; France.\\
$^{102}$Department of Physics, University of Massachusetts, Amherst MA; United States of America.\\
% ...
$^{116}$Department of Physics, New York University, New York NY; United States of America.\\
% ...
$^{119}$Homer L. Dodge Department of Physics and Astronomy, University of Oklahoma, Norman OK; United States of America.\\
% ...
$^{136}$$^{(a)}$Departamento de F\'isica, Pontificia Universidad Cat\'olica de Chile, Santiago;$^{(b)}$Millennium Institute for Subatomic physics at high energy frontier (SAPHIR), Santiago;$^{(c)}$Instituto de Investigaci\'on Multidisciplinario en Ciencia y Tecnolog\'ia, y Departamento de F\'isica, Universidad de La Serena;$^{(d)}$Universidad Andres Bello, Department of Physics, Santiago;$^{(e)}$Instituto de Alta Investigaci\'on, Universidad de Tarapac\'a, Arica;$^{(f)}$Departamento de F\'isica, Universidad T\'ecnica Federico Santa Mar\'ia, Valpara\'iso; Chile.\\
% ...
$^{149}$Department of Physics, Technion, Israel Institute of Technology, Haifa; Israel.\\
$^{150}$Raymond and Beverly Sackler School of Physics and Astronomy, Tel Aviv University, Tel Aviv; Israel.\\
% ...
$^{a}$ Also Affiliated with an institute covered by a cooperation agreement with CERN.\\
$^{b}$ Also at An-Najah National University, Nablus; Palestine.\\
$^{c}$ Also at Borough of Manhattan Community College, City University of New York, New York NY; United States of America.\\
$^{d}$ Also at Bruno Kessler Foundation, Trento; Italy.\\
$^{e}$ Also at Center for High Energy Physics, Peking University; China.\\
$^{f}$ Also at Centro Studi e Ricerche Enrico Fermi; Italy.\\
% ...

\end{flushleft}


\end{document}